\documentclass[twocolumn]{aastex631}

\pdfoutput=1 
\usepackage{amsmath,amstext}
\usepackage[T1]{fontenc}
\usepackage{apjfonts} 
\usepackage[figure,figure*]{hypcap}
\usepackage{multirow}
\usepackage{graphicx}
\usepackage{booktabs}
\graphicspath{{figs/}}

\begin{document}
\title{Modeling the GRB jet properties with 3D general relativistic simulations of magnetically arrested accretion flows} 

\correspondingauthor{Bestin James}
\email{bestin@cft.edu.pl}

\author[0000-0003-2088-7713]{Bestin James}
\author[0000-0002-1622-3036]{Agnieszka Janiuk}
\author[0000-0003-1449-0824]{Fatemeh Hossein Nouri}
\affiliation{Center for Theoretical Physics, Polish Academy of Sciences,
Al. Lotnik\'ow 32/46, 02-668 Warsaw, Poland}

\begin{abstract}

We investigate the dependence of the GRB jet structure and its evolution on the properties of the accreting torus in the central engine. Our models numerically evolve the accretion disk around a Kerr black hole using 3D general relativistic magnetohydrodynamic simulations. We use two different analytical hydrodynamical models of the accretion disk, based on the Fishbone-Moncrief and Chakrabarti solutions, as our initial states for the structure of the collapsar disk and the remnant after a binary neutron star merger, respectively. 
We impose poloidal magnetic fields of two different geometries upon the initial stable solutions. We study the formation and evolution of the magnetically arrested disk state and its effect on the properties of the emitted jet. The jets produced in our models are structured and have a relatively hollow core and reach higher Lorentz factors at an angle $\gtrsim 9{^\circ}$ from the axis. The jet in our short GRB model has an opening angle of up to $\sim 25^{\circ}$ while our long GRB engine produces a narrower jet, of up to $\sim 11^{\circ}$. 
We also study the time variability of the jets and provide an estimate of the minimum variability timescale in our models. The application of our models to the GRB jets in the binary neutron star post-merger system and to the ultra-relativistic jets launched from collapsing stars are briefly discussed.


\end{abstract}

\section{Introduction}\label{intro}

Gamma-ray bursts (GRBs) are transient phenomena observed in the high energy sky at cosmological distances. The emission of high energy photons is released at the jet photosphere and presents non-thermal spectral distribution \citep{Piran}. Their bi-modal duration distribution suggests separate classes of progenitors being responsible for short and long events \citep{1993ApJ...413L.101K}. The first class, long gamma ray bursts, have been identified with bright supernovae already in 1990s, e.g. for GRB 980425 \citep{1999A&AS..138..465G}. According to the collapsar model \citep{woosley1993} the jet emerges after the collapse of a massive, rotating star and formation of a black hole in its core. The accompanying explosion gives supernova-like signatures in the emission spectra, up to several weeks after the GRB. Magnetohydrodynamic (MHD) simulations have been performed for such a model of GRBs as early as \cite{ProgaCollapsar2003} and found that MHD effects alone are able to launch, accelerate, and sustain strong polar outflows.
The second class of bursts, whose duration is typically below couple of seconds, originates from compact binary merger events. Here the central compact object remnant is surrounded by an accretion disk that is created from remnant matter of a tidally disrupted neutron star \citep{2010CQGra..27k4002D}. As a result, a Poynting-flux dominated jet can be generated self-consistently as part of the magnetohydrodynamic processes in the merger \citep{rezzolla2011, ruiz2020}. A confirmation of this type of progenitor came recently with the discovery of gravitational wave event GW170817 associated with short GRB of duration 1.7 seconds in its prompt phase \citep{GW170817}. The accompanying observations showed clearly the multi-messenger characteristics of this object \citep{2021ARA&A..59..155M}.

GRBs are seen as relativistic jets pointing towards our line of sight, when observed from Earth. Relativistic jets are ubiquitous phenomena in many accreting black hole sources. 
It is widely assumed that the properties of the accretion inflow affect the properties of the jet.
The process which is responsible for driving accretion in magnetized disks is considered to be the magnetorotational instability (MRI) as described by \cite{BalbusHaw1991}.
The formation of the magnetically arrested disk (MAD) has been invoked recently to explain the properties of the jets observed (see \cite{AbramowiczFragile2013} and \cite{KoganMADreview2019} for a review). In this scenario a large scale bipolar field is accumulated around the central object due to the inward accretion of the plasma.  Such a large-scale field is unable to dissipate locally by the magnetic diffusivity, unlike a small-scale field, and cannot be absorbed if the central object is a black hole. The field thus accumulates in the innermost region of the accretion disk and results in the formation of a MAD \citep{Narayanetal2003}. 
The accumulation of magnetic flux which impedes with the accretion was noted even before in MHD simulations, which was later termed as the formation of a MAD \citep{ProgaBegelman2003, Igumenshchevetal2003}. In this situation, the interchange instability comes into play subsiding the MRI which already should have developed initially due to the turbulence in the magnetized accretion flow \citep{Proga2006}. 
The accretion further proceeds in a MAD state mainly due to the interchange instability.
The presence of large-scale bipolar fields in the accretion disks is often associated with the formation of astrophysical jets observed at various scales. Depending on the mass load, the magnetically driven jets can be classified into Poynting jets and hydromagnetic jets. The Poynting jets are naturally self-collimated and are powered either by the disks themselves or by the rotation of the black hole \citep{BZ1977, 2000ApJ...531L.111L}. The existence of the MAD state can be linked to the production of powerful Poynting jets \citep{Igumenshchev2008}. \cite{Tchekhovskoyetal2011} showed the formation of relativistic jet from a MAD state and the extraction of rotational energy of the black hole by the emitted jet according to the Blandford-Znajek mechanism. 
The other possible mechanism for jet launching in short GRBs is explained by neutrino-dominated accretion flows (NDAFs) \citep{Popham99,2002ApJ...579..706D,2004MNRAS.355..950J,2017Liu_Gu}. Since the post-merger disks are transparent for neutrinos, NDAFs are cooled continuously by neutrino emissions. This scenario proposes that some emitted neutrino energy can be transferred to a pair fireball through neutrino-antineutrino annihilations to generate the collimated jets along the axis perpendicular to the disk plane \citep{1991AcA....41..257P, Jaroszynski:1995mf, richers:15, Just:2016dba, Perego2017}.

The MAD models have been studied previously to explain the jet formation and time variability's dependencies on black hole's spin \citep{Narayan2021}, and formation of the blazar gamma-ray flares in active galactic nuclei (AGNs) and supermassive black holes at the galaxy centers such as M87 \citep{2018MNRAS.479.2534M,Chael2019}. \cite{Liska2020} investigated the effects of the initial magnetic field configuration and \cite{White2019} performed numerical convergence studies of the MADs.  Moreover, the MADs were applied in the context of the GRB observations explaining the variability of long GRB's luminosity during the prompt phase \citep{Lloyd-Ronning2016}, and constraining the magnetic field and black hole mass required to power Blandford-Znajek jets \citep{Lloyd-Ronning2019a}.

In this article, we aim to explain the jet properties of long and short GRB engines considering a magnetically arrested disk as the central engine. 
We use the method of \cite{Janiuketal2021} to determine the variability properties of these jets, extending now our previous work to the long-time, 3-dimensional simulation of MADs.
We investigate the properties of the relativistic jets produced by an accreting system around a Kerr black hole with two different hydrodynamical models of the accertion disk, using 3-dimensional general relativistic magnetohydrodynamic (GRMHD) simulations. We impose the initial magnetic field configurations such that the inner region of the disk builds up a substantial amount of poloidal flux in a short amount of time. 
Our intention was to achieve the MAD state rather faster and to study the dependence of the jet properties on such a central engine configuration. 

The article is organized in the following way. In Section \ref{models} we present the numerical setup and initial configuration of our models. In Section \ref{results} we describe the evolution of the disk, the formation of the MAD state, the properties of the resulting jet structure and some analysis of our results. The astrophysical implications of our models, the application of our results to the short and long GRBs and some further analysis are given in Section \ref{applications}. Finally we give a short summary and the conclusions in Section \ref{conclusions}.

\section{Numerical Setup and Models}\label{models}

\subsection{Code}
We use our implementation of the GRMHD code \texttt{HARM} \citep{Gammie2003, Noble2006, Sap2019} for evolving our models in a fixed Kerr metric. It is a conservative and shock capturing scheme for evolving the equations of GRMHD. The code follows the flow evolution by numerically solving the continuity, energy-momentum conservation, and induction equations in the GRMHD scheme 

\begin{equation}
    \nabla_\mu(\rho u^{\mu})= 0
\end{equation}    

\begin{equation}
    \nabla_\mu(T^{\mu\nu}) = 0
\end{equation}

\begin{equation}
    \nabla_\mu(u^\nu b^\mu - u^\mu b^\nu) = 0
\end{equation}

Here, $u^{\mu}$ is the four-velocity of the gas, $u$ is the internal energy, $\rho$ is the gas density, $p$ is the gas pressure, and $b^{\mu}$ is the magnetic four vector. The stress-energy tensor is comprised of the gas and electromagnetic parts:
     $T^{\mu\nu} = T^{\mu\nu}_{gas} + T^{\mu\nu}_{EM}$  where,

\begin{equation}
      T^{\mu\nu}_{gas} = \rho h u^\mu u^\nu + p g^{\mu\nu} = (\rho+u+p)u^{\mu}u^{\nu}+pg^{\mu\nu} ,
\end{equation}

\begin{equation}
    T^{\mu\nu}_{EM} = b^2 u^{\mu} u^{\nu}+ \frac{1}{2} b^2 g^{\mu \nu} - b^{\mu}b^{\nu} ,  
    b^{\mu} = u^{*}_{\nu} F^{\mu \nu}
\end{equation}

Here, $F$ is the Faraday tensor and in a force-free approximation, we have $E_{\nu} = u^{\nu}F^{\mu \nu} = 0$.
We adopt dimensionless units in the code, with $G = c = M = 1$ for our simulations. Thus the length in the code units is given by $r_g = GM/c^2$ and the time is given by $t_g = GM/c^3$, where M is the mass of the black hole. So our models can represent the central engines of both the short and long gamma ray bursts. 

The initial equilibrium torus state is prescribed in the Boyer-Lindquist coordinates (\cite{BL1967}) in the original solutions and they are transformed into Kerr-Schild (KS) coordinates in the code (see \cite{Weinberg1972-WEIGAC} \& \cite{2007arXiv0706.0622V}). The integration is done in the code in modified Kerr-Schild coordinates. So in the code, the KS radius $r$ has been replaced by a logarithmic radial coordinate $x^{[1]}$ such that $r = e^{x^{[1]}}$, the KS latitude $\theta$ has been replaced by $x^{[2]}$ such that $\theta = \pi x^{[2]} + \frac{(1-h)}{2}sin(2\pi x^{[2]})$ and the azimuthal angle $\phi$ remains the same, $\phi = x^{[3]}$. Here the parameter $h$ can be adjusted to concentrate the numerical resolution near to the mid-plane and we use a value of 0.5 for it in our models.

\subsection{Models}

The initial state of our models is assumed to be a pressure equilibrium torus, which is embedded in a poloidal magnetic field. The Kerr black hole will accrete matter onto it due to the development of the MRI in the disk and this will in turn affect the evolution of magnetic field. Our models have a black hole spin characterized by the Kerr parameter $a = 0.9$. It is in the range of equilibrium spin values estimated by \cite{GammieSpin2004} for the models with stellar mass black holes.
The simulations are run with a resolution of $288\times256\times128$ in $r$, $\theta$ and $\phi$ directions, respectively. We use the $\gamma-$ law equation of state
$p_g = (\gamma-1) u$ in all our models, where $p_g$ is the gas pressure, $\rho$ is the gas density and $u$ is the internal energy. We use a value of $4/3$ for $\gamma$.

For our first model (FM76), the initial state of the accreting torus is prescribed according to \cite{FM1976} (hereafter FM) who gave an analytic solution of a constant angular momentum steady state. We specify the initial size of the torus in geometrical units where the its inner edge $r_{in}$ is located at $6 r_g$ and the radius of pressure maximum $r_{max}$ is located at $13 r_g$. 
This solution corresponds to a constant specific angular momentum $l = 7.414$ in the torus.
For this model the outer edge of the disk is located around 60 $r_g$. We embed this initial torus configuration in a poloidal magnetic field, where the field lines follow the isocontours of density (see the top left panel in Fig.\ref{rho_B_fm}). For this field, the only non-zero component of the magnetic vector potential is given by 

\begin{equation}
A_{\phi} ({r,\theta}) = r^5 (\rho_{avg}/\rho_{max}) - 0.2 
\end{equation}

which has a dependence on the disk density structure as well as a power of radius. 
Here, $\rho_{max}$ is the initial density maximum inside the torus and $\rho_{avg}$ is the average of the density value from the adjacent cells in the computational grid. 
We choose this field configuration with the intention of eventually bringing a large poloidal flux to the vicinity of the black hole. We use the initial gas to magnetic pressure ratio, $\beta = p_{gas} /p_{mag}$, to normalize the magnetic field in the torus. 
Here $p_{gas} = (\gamma - 1)u_{max}$ and $p_{mag} = b_{max}^2/2$, where $u_{max}$ is the internal energy of the gas at the radius of maximum pressure. We normalize $\beta$ to a value 100 at the radius of maximum gas pressure, $r_{max}$ in the torus. 

Our second hydrodynamical model (Ch85) of the accreting torus uses the \cite{Chakrabarti1985} solution as the initial state. In this model the angular momentum has a power law relation along the radius with the von Zeipel parameter $\lambda = (l/\Omega)^{1/2}$, where $l$ denotes the specific angular momentum and $\Omega$ denotes the angular velocity. We adjust the size of the torus in the geometrical units using the parameters $r_{\rm in(vonZeipel)}$, which is the true inner edge of the disk in this model, with a value of $6 r_g$ and the radius of pressure maximum $r_{\rm max}$ with a value of $16.95 r_g$. We use different values here from that of the FM model, due to the different hydro-dynamical structure of the torus. The initial masses of the disks in both models are in the same range. The initial state of this model is shown in the left panel of the Figure \ref{rho_B_ch}. In this model the outer edge of the torus is located around $80 r_g$. 
In order to compare the size of the disk in both the models, we calculate the total mass of the disk at the initial time. 
Using the appropriate density unit scaling, for the physical models we are considering,
these values are estimated as $0.925 M_\odot$ and $0.132 M_\odot$ respectively for the FM and Chakrabarti solutions. In our scenario, these models represent separately the long and short GRB central engines. 
In our second model, we embed the initial torus configuration in a poloidal magnetic field. Here, the field is given as the magnetic field produced by a circular current. The only non-vanishing component of the vector potential in such a configuration is given by (e.g. in \cite{Jackson1998}):
\begin{eqnarray}
  A_{\phi}({r,\theta})= A_0 \frac{\left(2-k^2\right) K\left(k^2\right)-2 E\left(k^2\right)}{k\sqrt{4 R r \sin\theta}} \\
  k = \sqrt{\frac{4 R r \sin\theta} { r^2 + R^2 + 2 r R \sin\theta}} \nonumber
\end{eqnarray}
where $E,K$ are the complete elliptic functions and $A_0$ is a quantity which can be used to scale the magnetic field. In our model the radius of the circular wire $R$ is taken as radius of pressure maximum $r_{\rm max}$ of the disk.  We scale the magnetic field across the torus using the initial gas to magnetic pressure ratio, $\beta=p_{gas}/p_{mag}$. It has a maximum value of 480 inside the disk and has an average value of 100 within the disk. Therefore both our models are embedded in initial magnetic fields of comparable strength. The summary of initial parameters used in our models are given in Table \ref{table:model_parameters}.

\begin{table*}[htb]
\centering
\begin{tabular}{  p{4em} p{12em}  p{6em}  p{5em}  p{9em}  p{5em}  p{5em}  } 

  \hline
  Model & Initial magnetic field geometry & Kerr parameter ($a$) & Inner edge of the disk $r_{in}$ (in $r_g$) & Radius of pressure maximum $r_{max}$ (in $r_g$) & Mass of the disk in $M_{\odot}$ & Initial $\beta$ in the disk \\ 
  \hline
  1. FM76 & Poloidal field - following the disk density structure & 0.90 & 6.0 & 13.0 & 0.925 & 100 \\ 
  2. Ch85 & Poloidal field - due to a circular current at $r_{max}$ & 0.90 & 6.0 & 16.95 & 0.132 & 100 \\ 
  \hline
\end{tabular}
\caption{The summary of initial parameters used in our models.}
\label{table:model_parameters}
\end{table*}

In order to break the axisymmetry, we introduce a 2\% amplitude random perturbation (as similar to \cite{2018MNRAS.479.2534M}) to the internal energy such that $u = u_0(0.98+0.1X)$ where, $u_0$ is the equilibrium internal energy in our models and $X$ is a random number in the range $0 \leq X \leq 1$. This perturbation breaks the initial axisymmetry of the system and helps in the development of non-axisymmetric modes.

\section{Results}\label{results}

\subsection{Initial configuration and evolution of the magnetized tori} \label{torus_evolution}

\begin{figure*}[htb]
\centering
\includegraphics[width=1\textwidth]{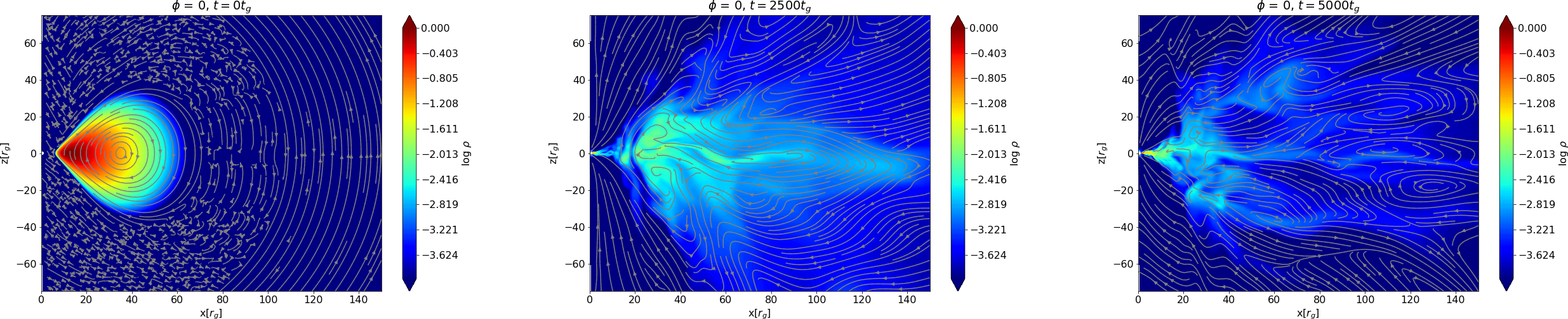}
\includegraphics[width=1\textwidth]{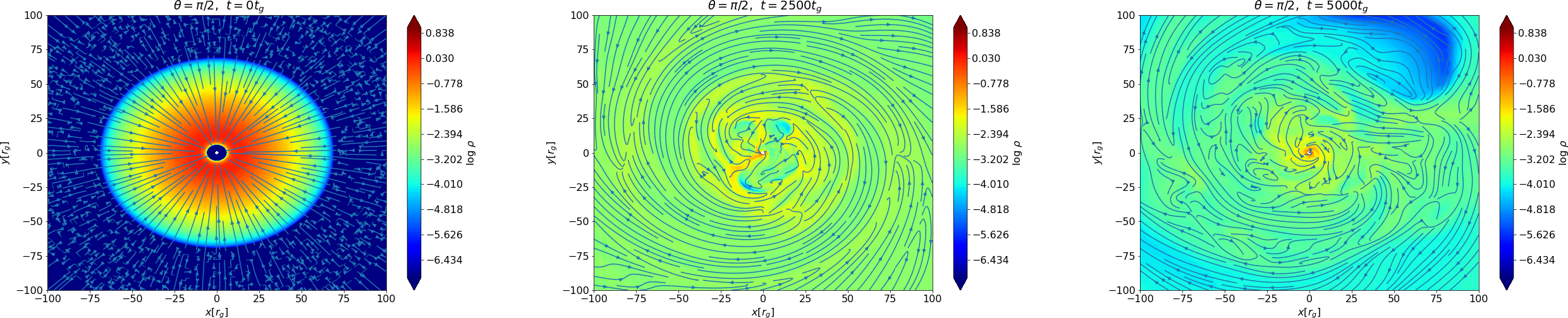}
\caption{Snapshots of torus density structure along a poloidal plane ($\phi = 0$ slice) (top panel) and along the equatorial plane ($\theta = \pi/2$ slice) (bottom panel) with streamlines of magnetic field at t= 0 (left), 2500 (middle) and 5000 $t_g$ (right), for the model with initial $\beta = 100$ and Fishbone-Moncrief configuration with Kerr parameter $a=0.90$.
}
\label{rho_B_fm}
\end{figure*}

\begin{figure*}[htb]
\centering
\includegraphics[width=1\textwidth]{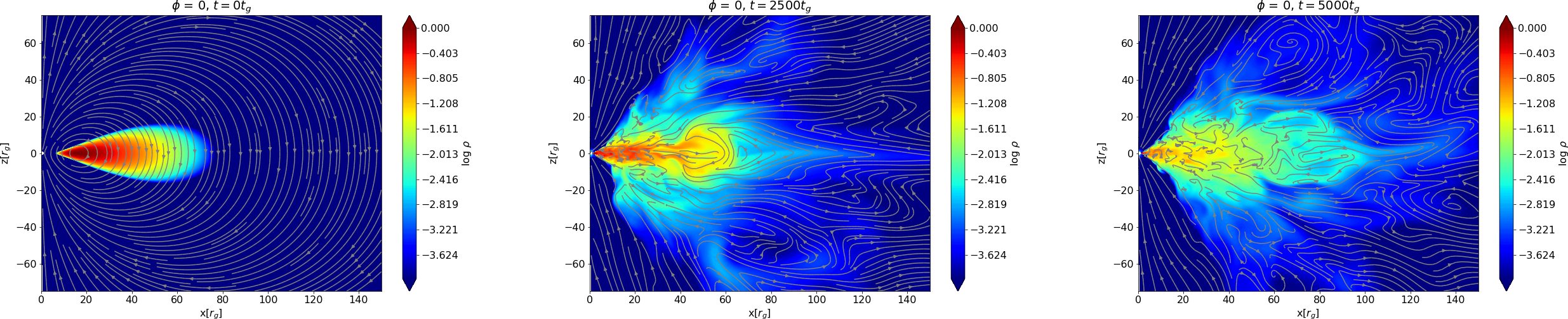}
\includegraphics[width=1\textwidth]{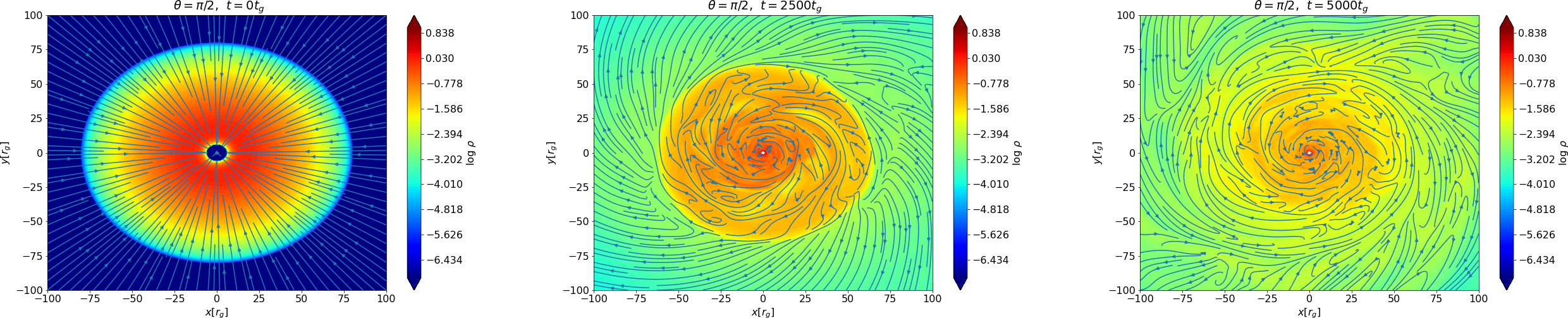}
\caption{Snapshots of torus density structure along a poloidal plane ($\phi = 0$ slice) (top panel) and along the equatorial plane ($\theta = \pi/2$ slice) (bottom panel) with streamlines of magnetic field at t= 0 (left), 2500 (middle) and 5000 $t_g$ (right), for the model with initial average $\beta = 100$ and Chakrabarti configuration with Kerr parameter $a=0.90$.
}
\label{rho_B_ch}
\end{figure*}

We start our simulations with a stable equilibrium analytic torus solution, as described in the previous section. The imposed magnetic field on the initial stable configuration causes turbulence inside the plasma and results in the MRI. This acts as a mechanism which transports angular momentum outwards from the torus. Thus the matter starts to accrete into the black hole. The geometry of the imposed magnetic field has some pronounced effect on the further evolution of the system. In our first model, the imposed field is based on the magnetic vector potential which is dependent on the fifth power of the radius. So the poloidal field is smoothed out over a large distance and it takes a certain time for the plasma to bring the flux nearer to the black hole horizon. Such a configuration is plausible for collapsar central engines, where the core collapse leads to magnetic flux compression (e.g. \cite{Burrowsetal2007magdrivSupernova}, see also \cite{Tchekhovskoyetal2011}). On the other hand in our second model, the initial poloidal field is already strong near the horizon since it has the maximum value at the radius of pressure maximum ($r_{max}$) of the disk. Such a configuration is adequate to describe the magnetic field in the post-merger remnant torus (see \cite{Paschalidisetal2013}).

The evolved states of the flow after some time has elapsed, illustrated by the density structure, are in the middle and right panels of the Figures \ref{rho_B_fm} and \ref{rho_B_ch}, at times t = 2500 $t_g$ and t = 5000 $t_g$ respectively. The over-plotted lines show the magnetic field. It can be noticed that the initial conditions have relaxed and the accretion has started due to the action of the magnetic field. 
We observe a couple of azimuthal modes of the Rayleigh-Taylor instability developing in the turbulent region of the inner disk. Hence, the accretion is steadily driven and proceeds despite the formation of a magnetic barrier.
When we plot the streamlines of velocity similarly at the same time instances, 
the velocity field lines loosely follow the structure of the density distribution inside the disk and along the equatorial region we can see the matter going in to the black hole. In the regions away from the disk, we observe outflows. In the poloidal slices for the evolved stages, we observe that the velocity streamlines always point outward near to the polar axis from the black hole which show the direction of jet flows in the model.

\begin{figure*}
    \centering
    \includegraphics[width=0.47\textwidth]{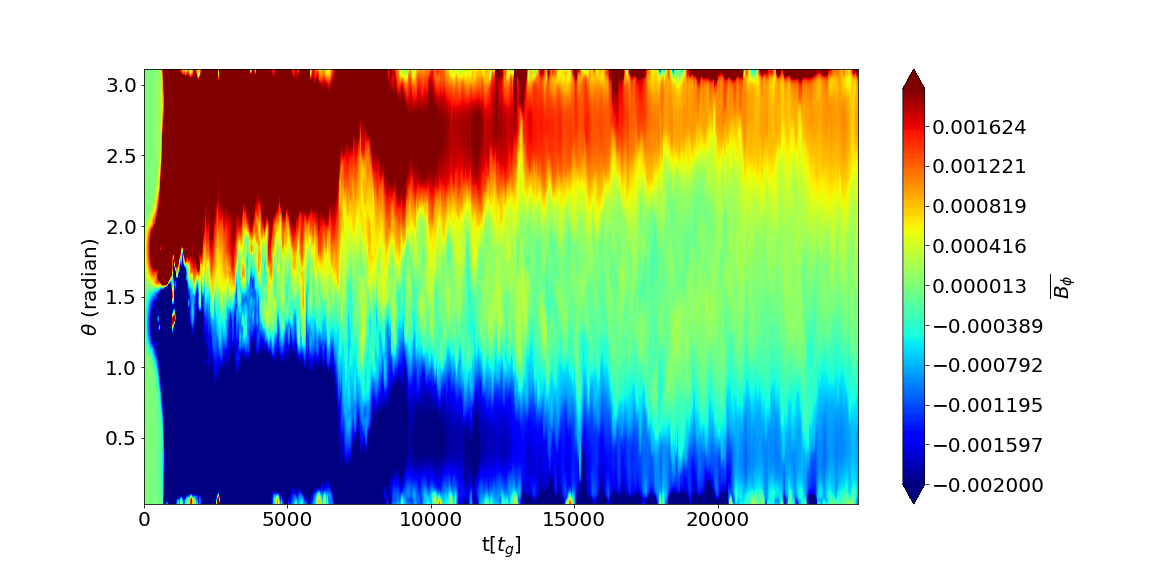}
    \includegraphics[width=0.47\textwidth]{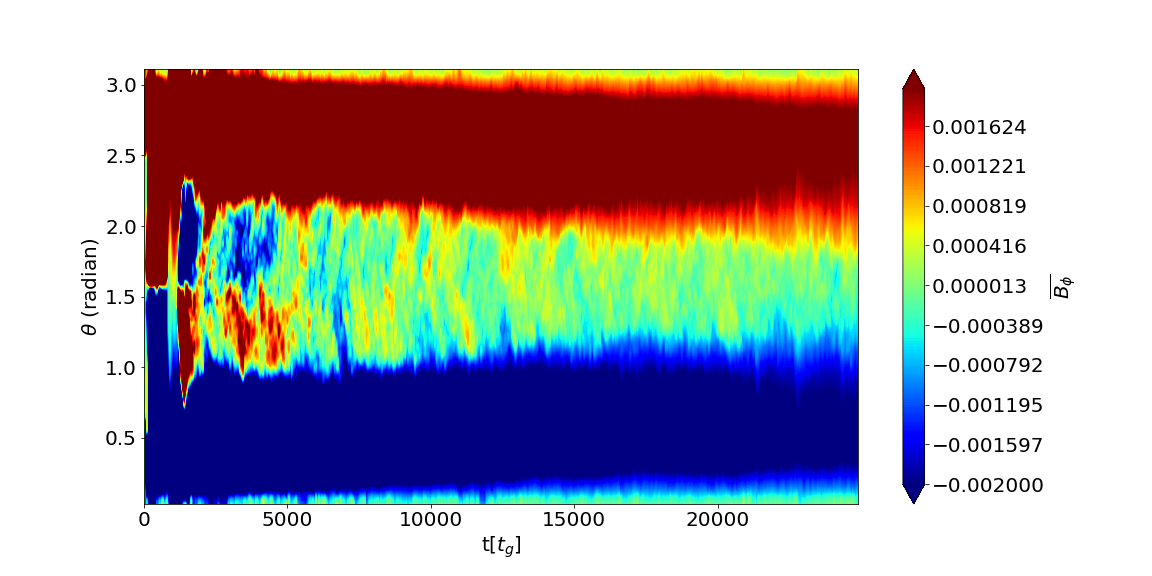}
    \caption{Butterfly diagram: showing the spatio-temporal evolution of the azimuthally averaged toroidal field ${\bar{B_{\Phi}}}$(r = 10$r_g$, $\theta$, t) in our models with (a) FM initial disk and (b) Chakrabarti initial disk configuration. 
    }
    \label{fig:butterfly}
\end{figure*}

\begin{figure*}
\centering
\includegraphics[width=0.47\textwidth]{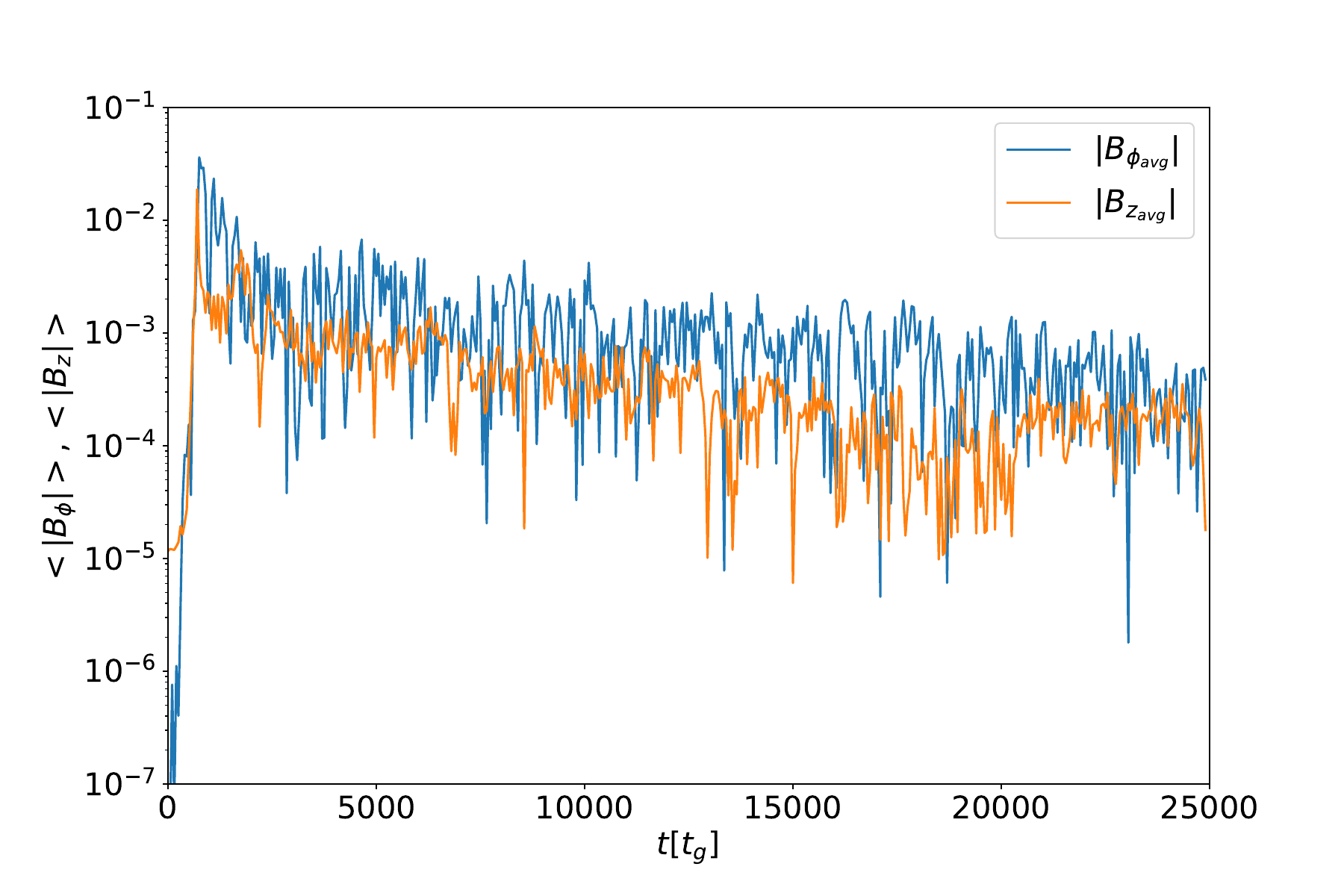}
\includegraphics[width=0.47\textwidth]{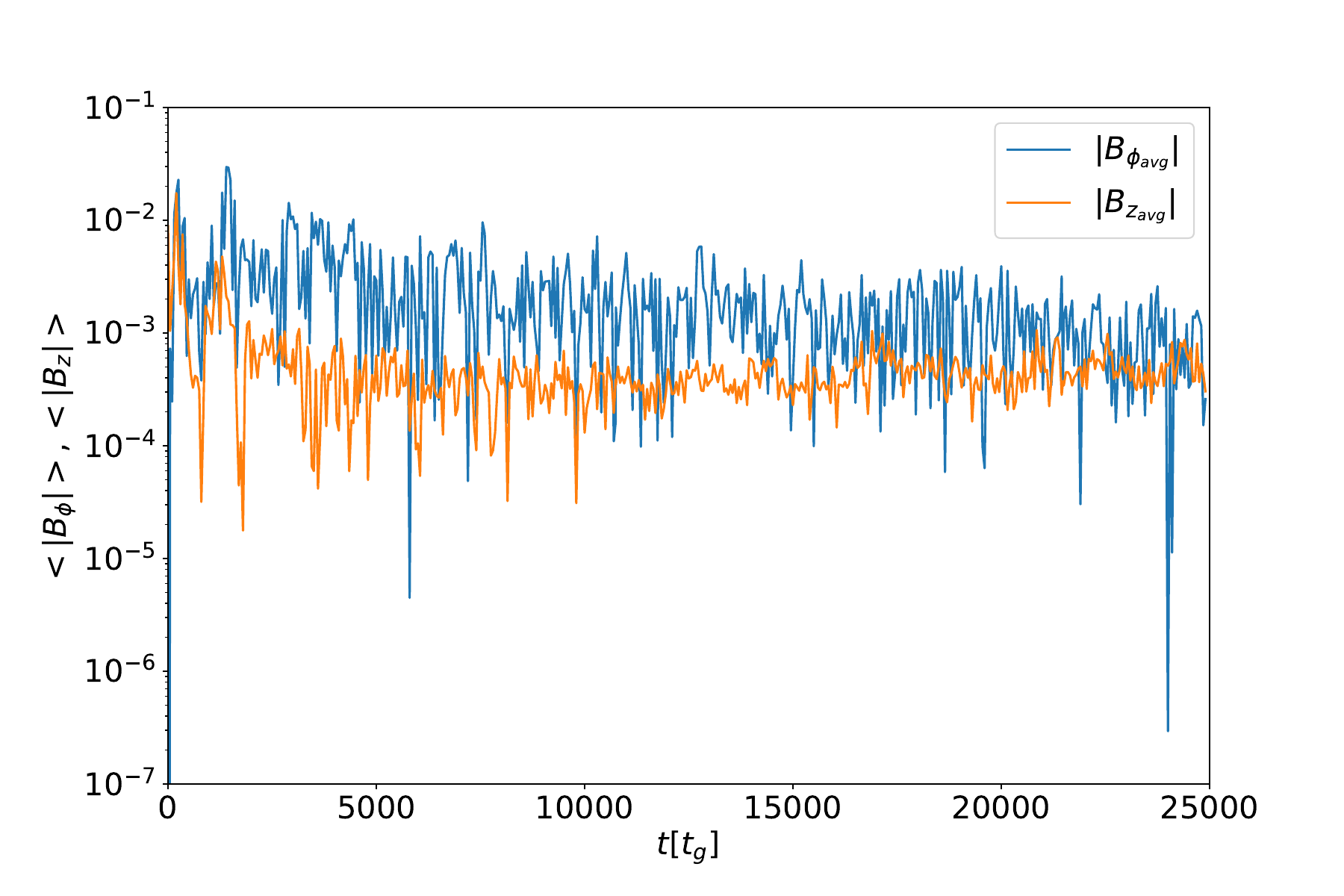}
\caption{Strengths of the toroidal and poloidal components of the magnetic field
at the equator ($\theta = \pi/2$) averaged over $r_{hor} \leq r \leq 10r_g$ and $0 \leq \phi \leq 2\pi$, with time (a) for model with initial Fishbone-Moncrief configuration (FM76) and (b) with initial Chakrabarti configuration (Ch85).
}
\label{Bav_t}
\end{figure*}

In Figure \ref{fig:butterfly} we show the evolution of the azimuthally averaged toroidal component of the magnetic field and its expansion with the polar angle ($\theta$) over time.
Initially, there is no magnetic field in the poles in either of the models, while the strong toroidal field is generated at the equatorial plane, in the accreting torus due to its differential rotation. The field is transported to the black hole horizon with the MRI turbulence and is present there shortly after the beginning of the simulation. The average shown in the plots is taken at $\phi=(0,2\pi)$ and at a radius of 10 $r_g$. 
With time, the toroidal field develops also at polar regions, where it is being wind-up by the rotation of the black hole. It is also transported from the accretion disk towards the intermediate latitudes, via magnetic buoyancy, but in these regions the strength of the field is smaller than at the poles.
In both our models we observe strips showing the periodic changes of the toroidal field component towards the polar regions. 
They seem to be anticorrelated with the pulses of poloidal magnetic field, depicted in Fig. \ref{Bav_t}, discussed below.
In the second model, we notice also the field reversal in the regions near to the equator, which occurs predominantly at the initial period of the simulation ranging from time 1500 $t_g$ to 6000 $t_g$.
This is not noticed in our FM model evolution, however \cite{2018MNRAS.479.2534M} noted the field reversals also in their FM simulations, albeit with twice smaller resolution than in our models.

In Figure \ref{Bav_t} we show the time evolution of the poloidal and toroidal components of the magnetic field in our models. The strength of the field components is calculated at the equator ($\theta = \pi/2$) by averaging it over the whole range of $\phi$ ($0 \leq \phi \leq 2\pi$) and also averaged for the radius in the range $r_{hor} \leq r \leq 10r_g$, where the magnetic field is mostly concentrated. The plots show that the initially imposed poloidal field gets amplified with time due to the turbulence in the plasma in both the models. It can also be observed that the torodial component is initially zero but reaches a considerable strength (up to an order of $\sim 10^{-2}$ in code units) over time due to the winding by the black hole rotation. The strength of the developed toroidal component remains higher in the disk than the poloidal counterpart for most of the simulation.  
The jet is launched thanks to the development of the toroidal magnetic field, which also helps in collimation of the jet base, in addition to the ram pressure acting at larger scales.

\begin{figure*}
\centering
\includegraphics[width=0.47\textwidth]{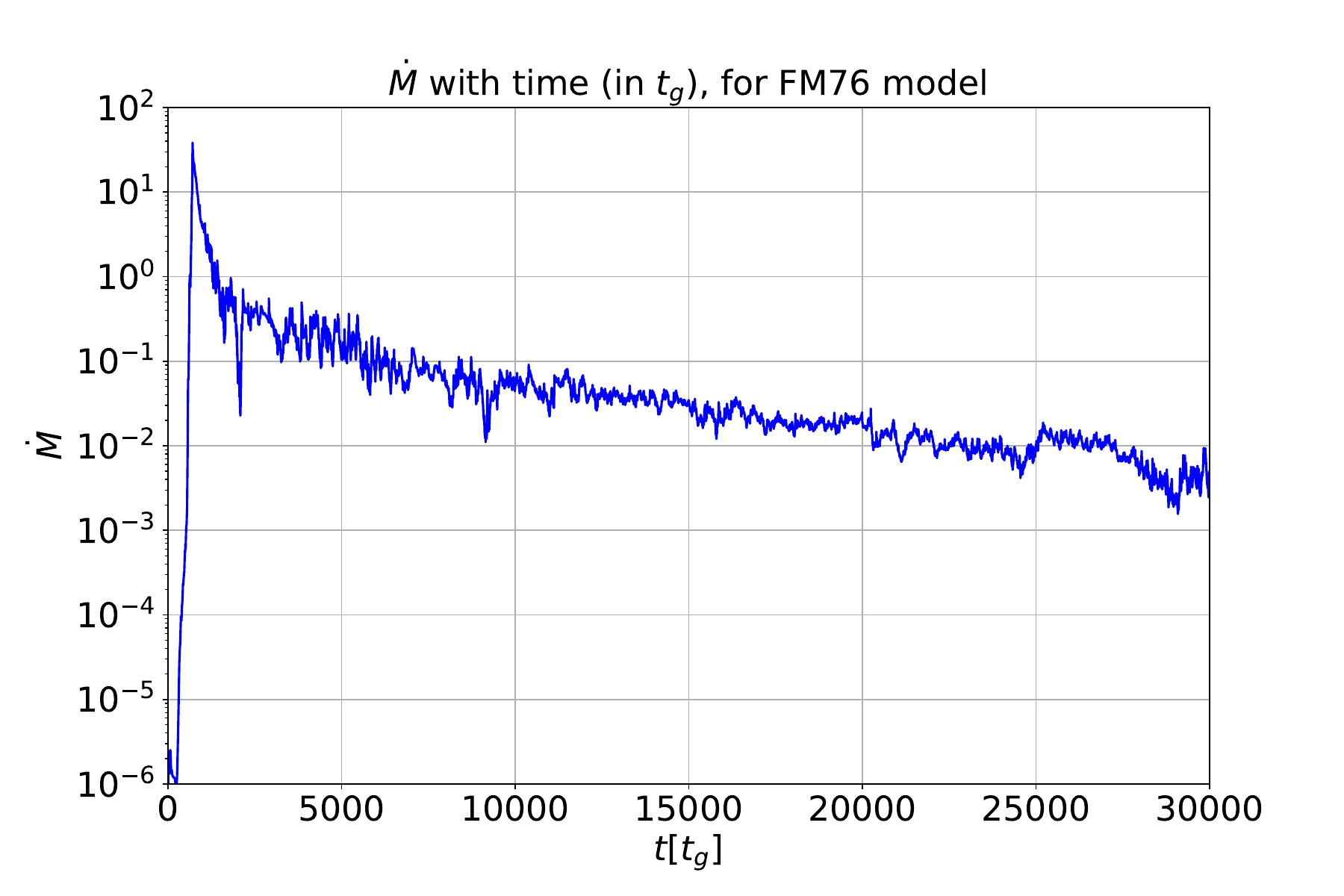}
\includegraphics[width=0.47\textwidth]{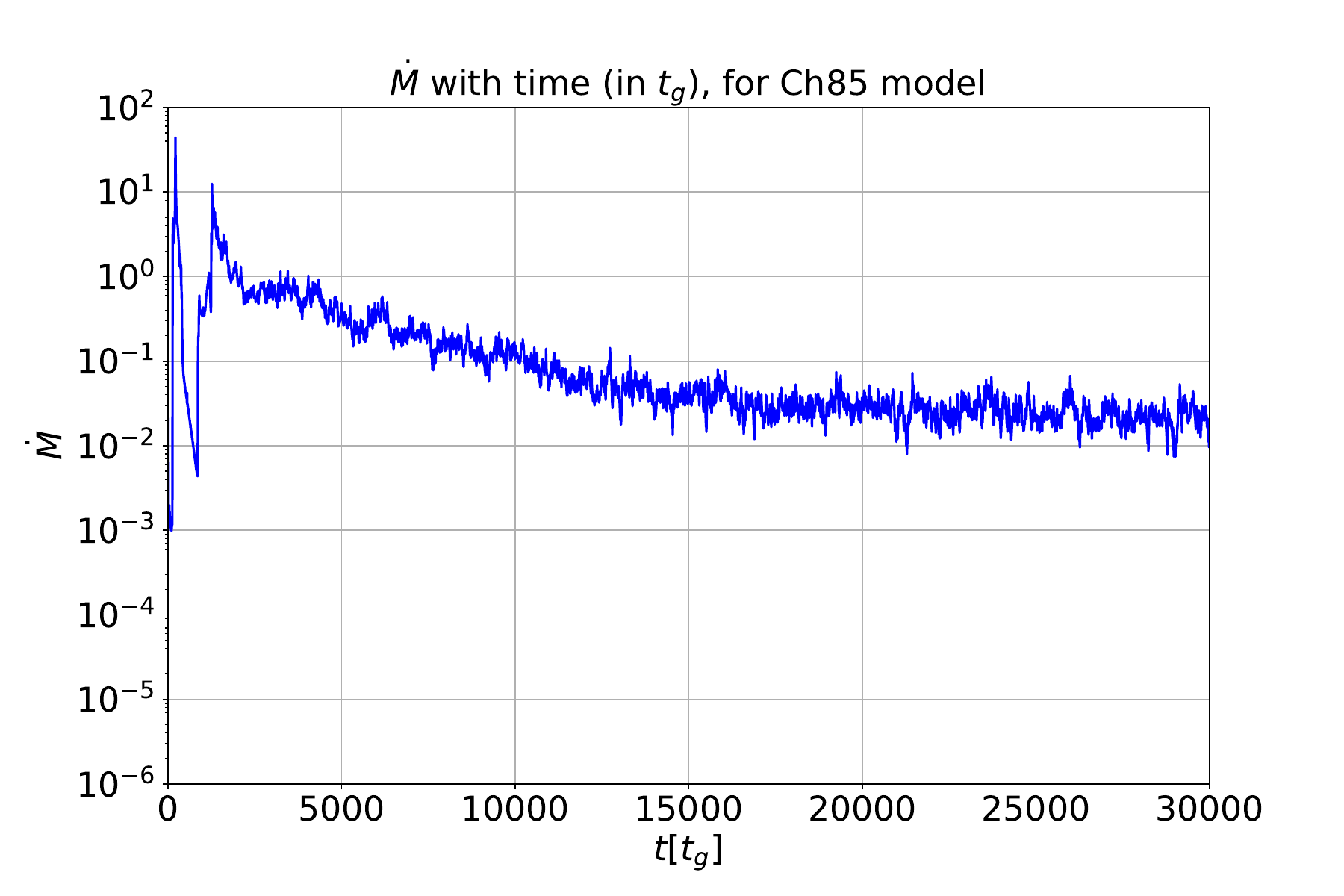}
\caption{Mass accretion rate at the event horizon ($\approx 1.4 r_g$) with time (in $t_g$) (a) for model with initial Fishbone-Moncrief configuration (FM76) and (b) with initial Chakrabarti configuration (Ch85).
}
\label{mdot_t}
\end{figure*}

In Figure \ref{mdot_t} we show the mass accretion rate near the black hole horizon as a function of time for both our models. The accretion starts after a short amount of time due to the development of the magnetic turbulence. This brings in plasma to the black hole horizon along with the magnetic flux. The sudden increase in the accretion rate in the beginning of the simulation can be attributed to the development of the magnetic turbulence which begins to bring the matter in the disk to the horizon and relaxation of the initial stationary conditions. But when a substantial amount of magnetic flux is built up near the horizon, the accretion significantly reduces shortly afterwards which can be seen as the sudden drop in the accretion rate after around 1500 $t_g$. This trend is observed in both our models but is more pronounced in the second one due the specific geometry of the initially applied field. This is connected to the formation of the magnetically arrested disk. Thus our second model reaches the magnetically arrested state much quicker than the first one. The mass accretion proceeds and sustains in time due to the plasma instabilities further developed in the disk.

\subsection{Formation of the magnetically arrested disk}

In our simulations the accretion of matter onto the black hole is initiated due to the MRI and proceeds further due to the sustained turbulence in the plasma. Depending on the initial geometry and strength of the magnetic field, it is probable that the accreting material brings a substantial amount of poloidal flux to the vicinity of the black hole over time. In such cases after a certain time, the strength of the magnetic field threading the black hole horizon increases considerably. A strong poloidal magnetic field developed in such a way will impede with the accretion and push away the plasma coming into the black hole. Thus the smooth flow of matter along the equatorial axis is halted and the further accretion proceeds mainly in short episodes due to the interchange instability developed in the plasma afterwards. Such a state of the accreting torus is often called as the magnetically arrested disk (MAD) state. We observe the MAD state in both of our models with full 3D simulations. To parameterize the amount magnetic flux on the black horizon as compared to the inward flow of matter, we calculate the normalized and averaged magnetic flux threading each hemisphere of the black hole horizon. It is computed as

\begin{equation}
    \phi_{BH}(t) = \frac{1}{2\sqrt{\dot{M}}} {\int_{\theta} \int_{\phi} \mid{B^{r}(r_{H},t)}\mid dA_{\theta \phi}}
    \label{eq:phi_BH}
\end{equation}

\begin{figure*}
\includegraphics[width=0.49\textwidth]{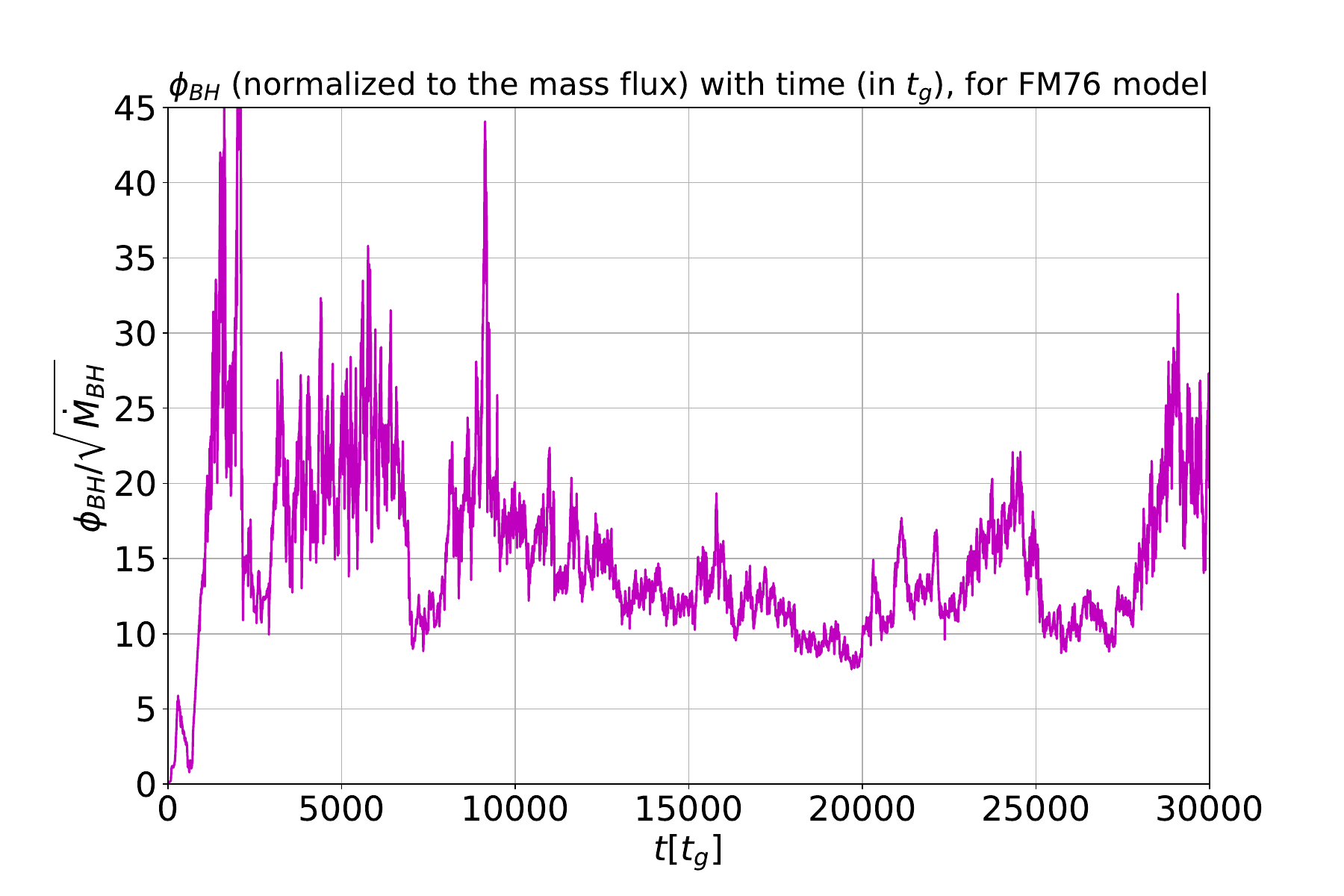}
\includegraphics[width=0.49\textwidth]{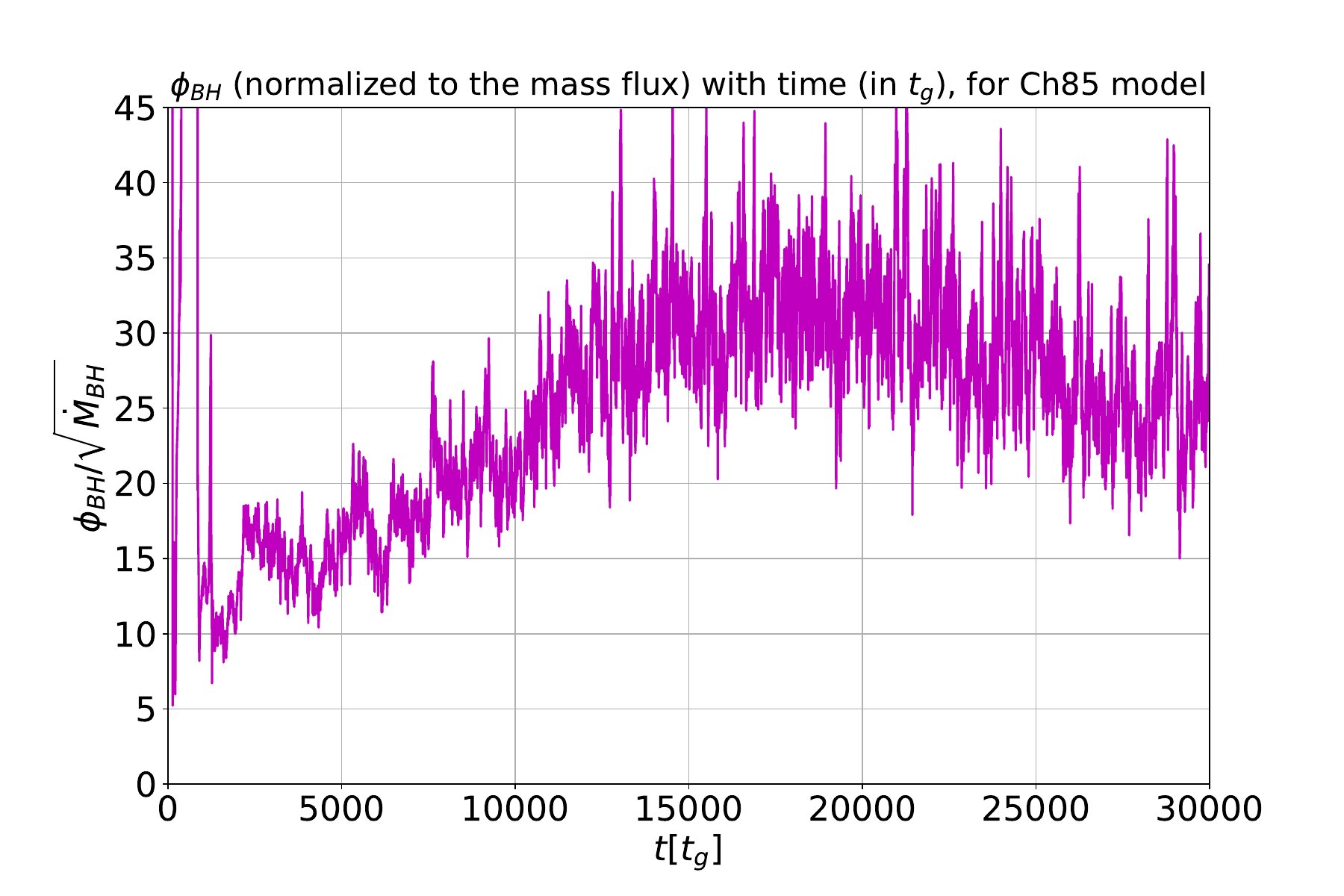}
\caption{Time evolution of the magnetic flux on the black hole horizon normalized to the mass flux (a) with initial Fishbone-Moncrief configuration (FM76) and (b) with initial Chakrabarti configuration (Ch85).
}
\label{fig:phi_BH}
\end{figure*}

The magnetic flux threading the black hole horizon, normalized to the mass flux, is shown in Figure \ref{fig:phi_BH} for both our models
(In our code we use the Gaussian units and so a factor of $4\pi$ is not included in this flux). It can be seen that our models have dynamically evolving magnetic fields in the inner disk region. In the model with FM initial configuration, we notice from the plot that the magnetic flux gets built up as the accretion proceeds, even though the initially applied field is not extremely strong, quantified by the plasma beta value as high as 100 inside the disk. After around 1000 dynamical times, the normalized magnetic flux reaches the highest value of up to 50. The mass inflow is thus hindered with due to the building up of the field and the disk reaches a magnetically arrested state. The mass accretion rate considerably reduces from above $10$ to the range of $10^{-1}$ to $10^{-3}$ (in code units) after such a state is achieved (see Figure \ref{mdot_t}). This normalized magnetic flux on the horizon always remain 10 times or more larger than the mass flux as the model further evolves and the disk stays in a magnetically arrested configuration. 
In the model with the Chakrabarti initial state, we see such an effect more immediately after the accretion starts. The normalized magnetic flux threading the black hole horizon reaches the value of upto 120, as compared to the mass flux. This higher value as compared to the previous model can be attributed to the different initial magnetic field configuration we use in this model, even though the strengths are comparable. But here also, the accretion proceeds afterwards due to the instabilities developed in the plasma afterwards, especially due to the interchange instability. The magnetic flux normalized to the mass flux at the horizon remains higher than 20 in this model for most part of the simulation and has a more dynamic nature. The time averaged values of this magnetic flux at the horizon are 15.33 for the FM76 model and 25.49 for the Ch85 model, respectively.

\subsection{Jet power and energetics}

We use the jet energetics parameter $\mu$ to estimate the Lorentz factor at infinity, assuming all energy is transformed to the baryon bulk kinetic, as shown by \cite{Vlahakis2003}. It is defined as: 
\begin{equation}
    \mu = - \frac{T^r_t}{\rho u^r}
    \label{eq:mu}
\end{equation}

Here, $T^r_t$ is the energy component of the energy-momentum tensor which comprises matter and electromagnetic parts, $\rho$ is the gas density and $u^r$ is the radial velocity. So it is the total plasma energy flux normalized to the mass flux. 
We also use the magnetization parameter $\sigma$ to estimate the degree of magnetization in the jet funnel region:

\begin{equation}
    \sigma = \frac{(T_{EM})^ r_t}{(T_{gas})^ r_t}
    \label{eq:sigma}
\end{equation}

\begin{figure*}
    \centering
    \includegraphics[width=0.24\textwidth]{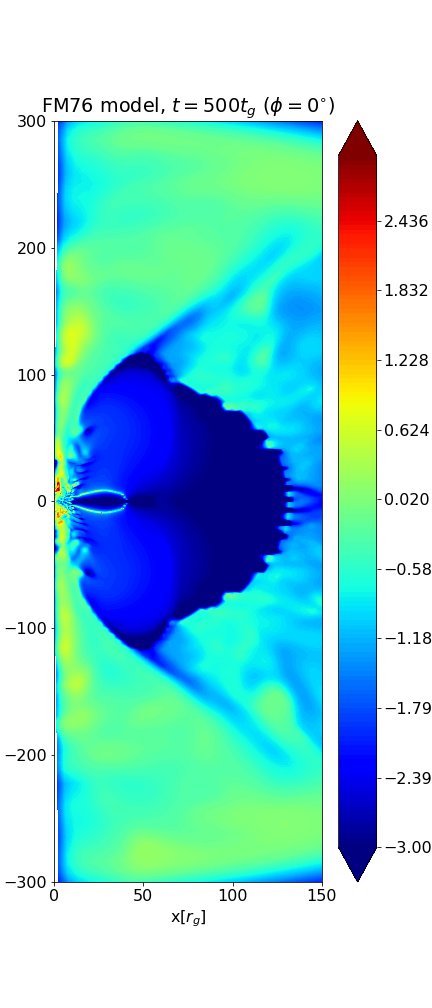}
    \includegraphics[width=0.24\textwidth]{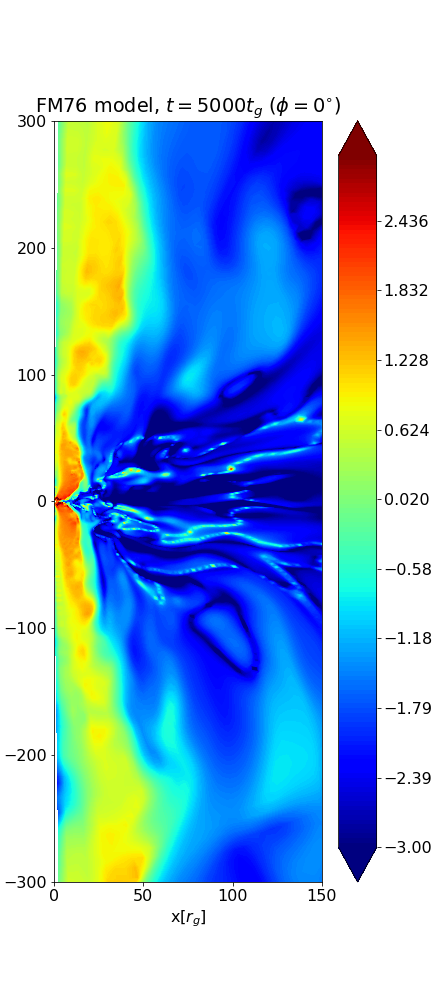}
    \includegraphics[width=0.24\textwidth]{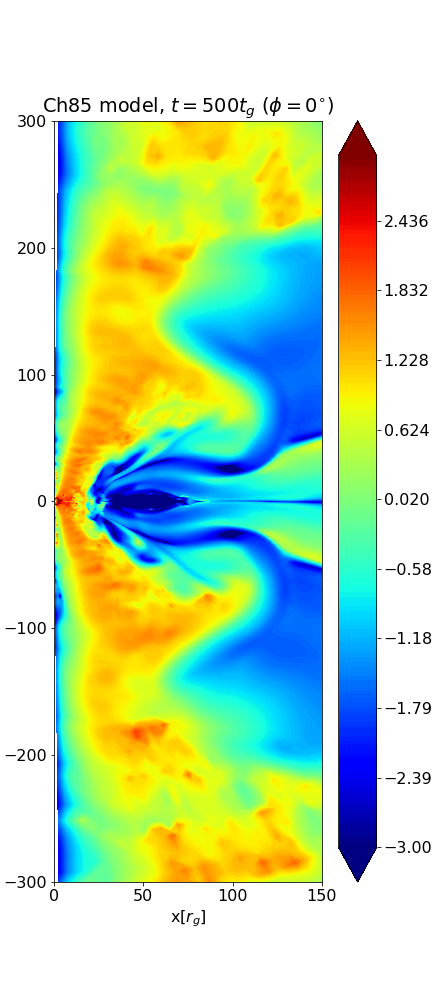}
    \includegraphics[width=0.24\textwidth]{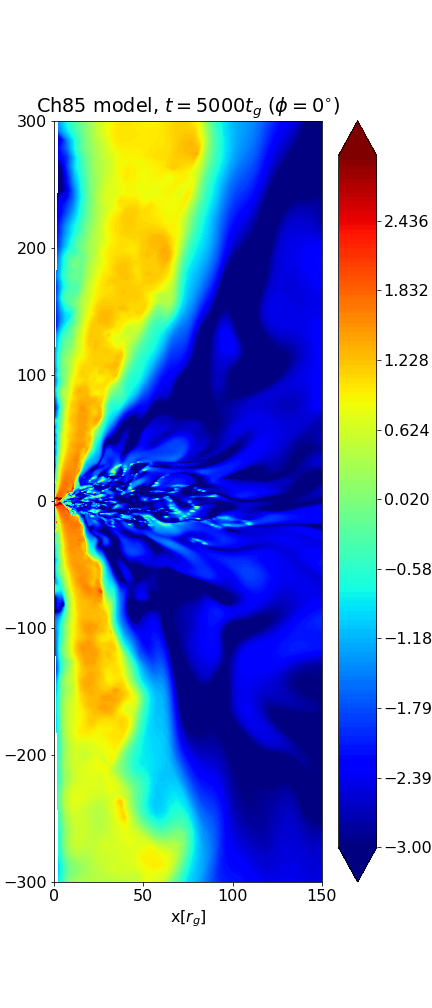}
    \caption{2D contours of jet magnetization parameter $\sigma$ in log scale (see Equation \ref{eq:sigma}) for (a) FM76 and (b) Ch85 models at time $t = 500 t_g$ (at the very beginning of the jet launching) and $t = 5000 t_g$ (at an evolved time) respectively. The plots show that the jet funnel region is magnetized from the initial times of jet launching.}
    \label{fig:sigma_2d}
\end{figure*}

In Figure \ref{fig:sigma_2d}, we plot the logarithmic contours of $\sigma$ at time snapshots $t = 500 ~t_g$ and $5000 ~t_g$, that is approximately at the jet launching phase and at an evolved time. The plots show that the jet funnel region is significantly magnetized already at the initial launch phase.  Especially in the Ch85 model, the maximum magnetisation is reaching $\log \sigma=2.5$, at the base of the jet. This high magnetisation persists at later times, and also develops eventually in the FM76 model, which was initially slightly less magnetised in the jet region.

\begin{figure*}
    \centering
    \includegraphics[width=0.48\textwidth]{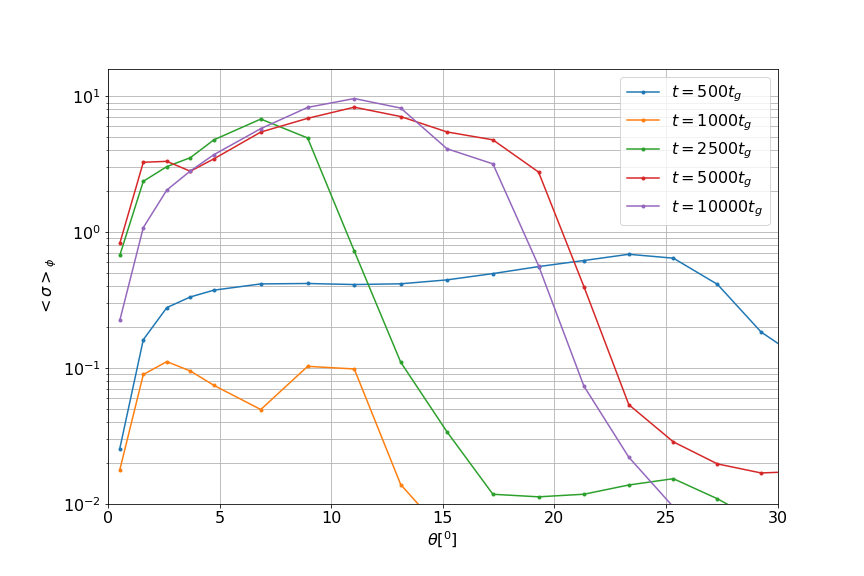}
    \includegraphics[width=0.48\textwidth]{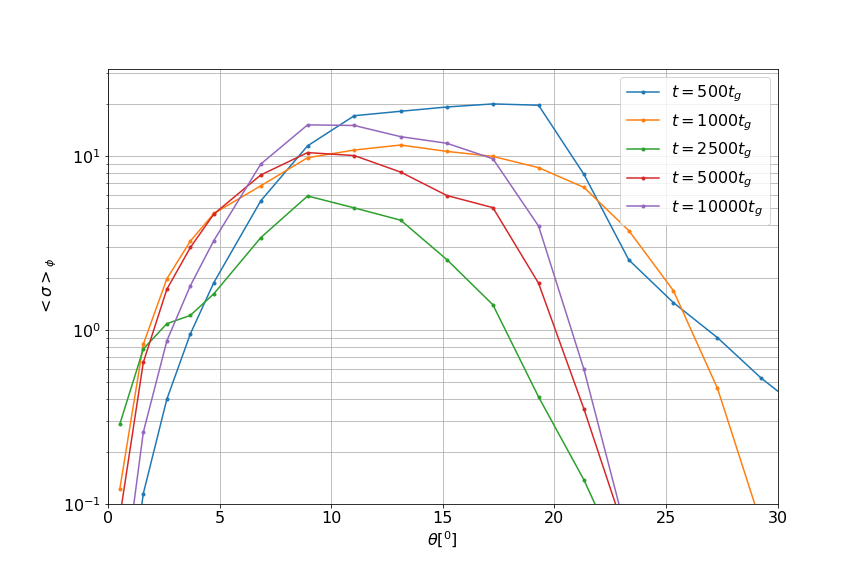}
    \caption{The profile of jet magnetization parameter $\sigma$ (see Equation \ref{eq:sigma}) for (a) FM76 model and (b) Ch85 model, computed at $r = 150~r_g$, with the polar angle $\theta$ at different time instances. The plots show that the magnetization is negligible along the polar axis at the initial time but increases fast in the jet funnel region as the jet launches. This shows that the jet is magnetized from the launching phase.}
    \label{fig:sigma_profile}
\end{figure*}

Figure \ref{fig:sigma_profile} shows the azimuthally-averaged, angular profiles of the magnetization parameter, 
taken at different time instances of evolution. It can be seen that the initially negligible magnetization increases eventually towards the jet walls, and the launching phase starts in both models around 500 - 1000 $t_g$. The level of magnetization fluctuates slightly in the jet funnel region due to the episodic accretion events.

\begin{figure*}[htb]
\centering
\includegraphics[width=0.48\textwidth]{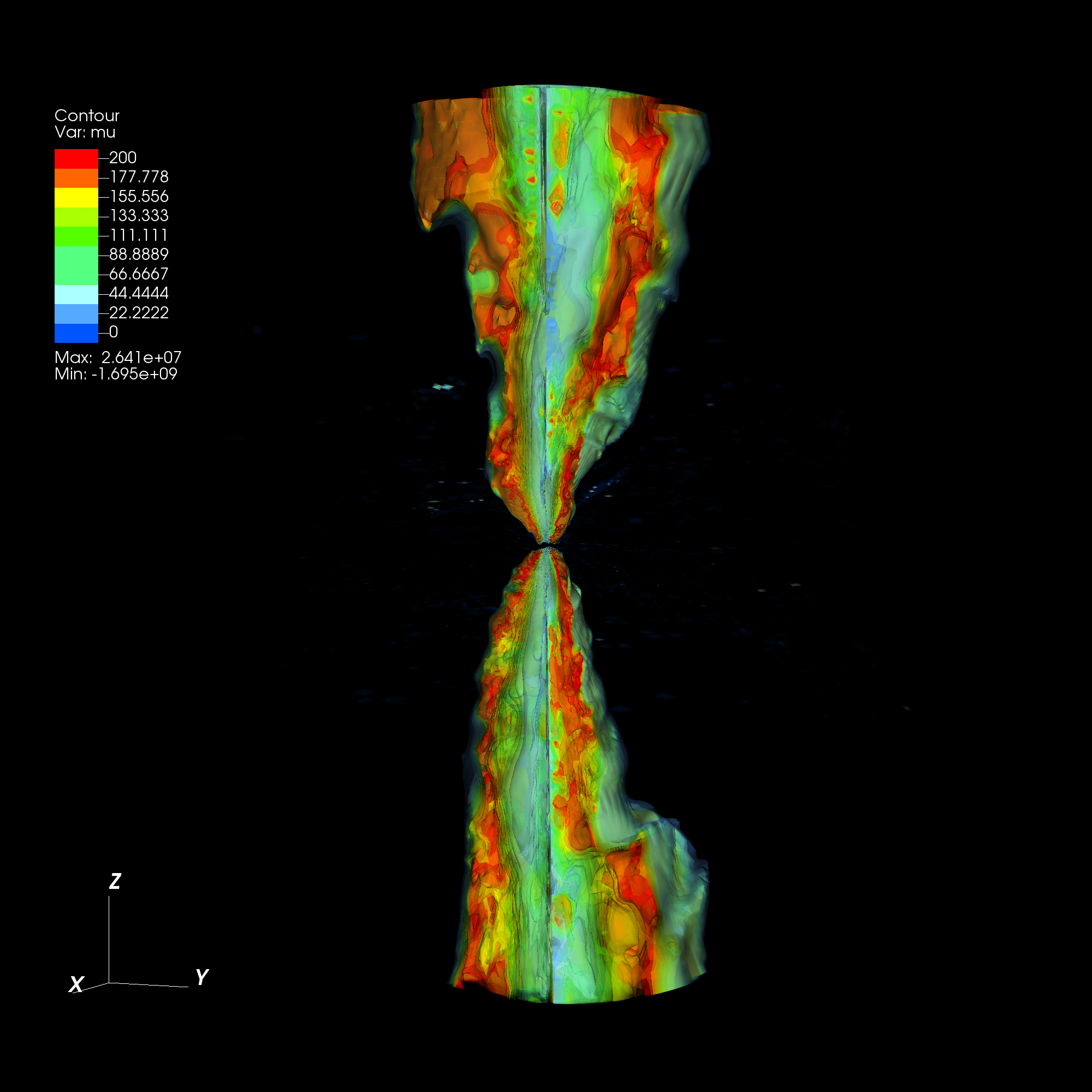}
\includegraphics[width=0.48\textwidth]{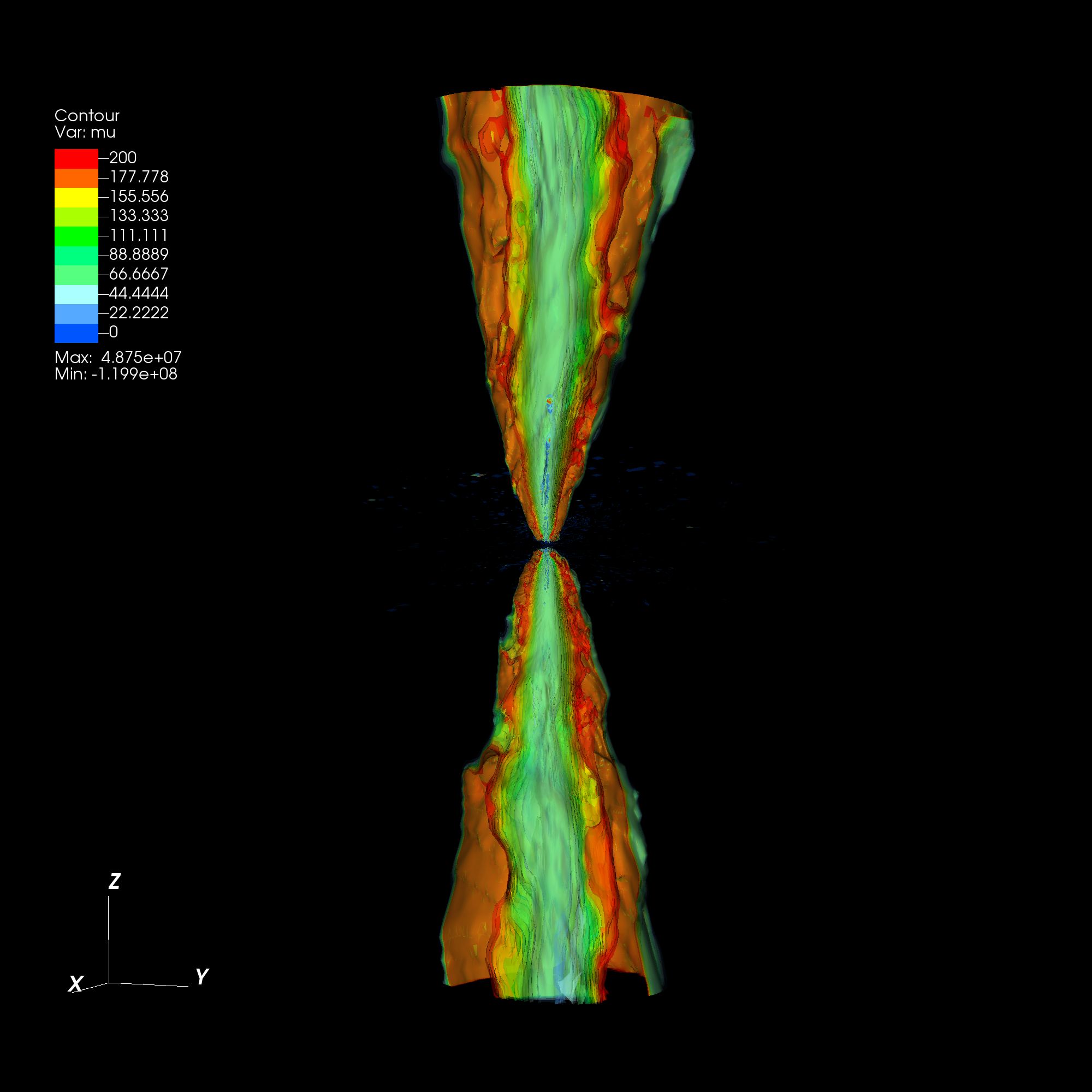}

\caption{3D jet structure at time t = 5000 $t_g$ for (a) Fishbone-Moncrief (FM76) model (left panel) and (b) Chakrabarti (Ch85) model (right panel). The plots show the contours of the energetics parameter defined as $\mu$ (see Equation \ref{eq:mu}) up to a radius of 200 $r_g$ in both directions. The plots are clipped along the YZ-plane to show the inner structure of the jet.
}
\label{fig_mu_2d_fm}
\end{figure*}

Figure \ref{fig_mu_2d_fm} shows the structure of the jet by the distribution of the $\mu$-parameter at 5000 $t_g$ for both the models. We choose two locations along the jet direction and estimate the value of the jet energetic parameter $\mu$ at these locations. The chosen location 1 is at $r = 150 r_g$, $\theta = 5 ^{\circ}$ which is in the inner region of the jet close the polar axis while the location 2 is at $r = 150 r_g$, $\theta = 10^{\circ}$ which is towards the outer region of the jet. The values we get from these locations are averaged over the whole toroidal angle $\phi$.
We calculate the Lorentz factor of the observed jets from our models as the time average of $\mu$ (after the jet has launched) at these chosen locations. The Lorentz factor averaged from the two locations for the FM model is 97.71 and for the Chakrabarti model is 131.85. So both the jets are accelerated to high relativistic velocities. The higher Lorentz factor in the second model may be attributed to the different initial structure of the magnetic field, which extends beyond the disk.

The time variability of the jet referring to the jet energetic parameter $\mu$ is discussed in the next subsection.

In order to estimate the Blandford-Znajek luminosity of the emitted jet, we compute the radial energy flux by (see \cite{McKinney2004})

\begin{equation}
\dot{E} \equiv \int_{0}^{2\pi}\int_{0}^{\pi} d\theta d\phi \sqrt{-g} F_E
\label{eq:p_jet}
\end{equation}

where $F_E \equiv -T^r_t$.  This can be further subdivided into matter $F_E^{(MA)}$ and electromagnetic $F_E^{(EM)}$ parts. The electromagnetic part, only which we consider in our computation above, is given by (see \cite{McKinney2012MagChoked})

\begin{equation}
{T^{(EM)}}^\mu_\nu = b^2 u^\mu u_\nu + p_b \delta^\mu_\nu - b^\mu b_\nu
\label{eq:Tr_t_EM}
\end{equation}

\begin{figure*}
\centering
\includegraphics[width=0.47\textwidth]{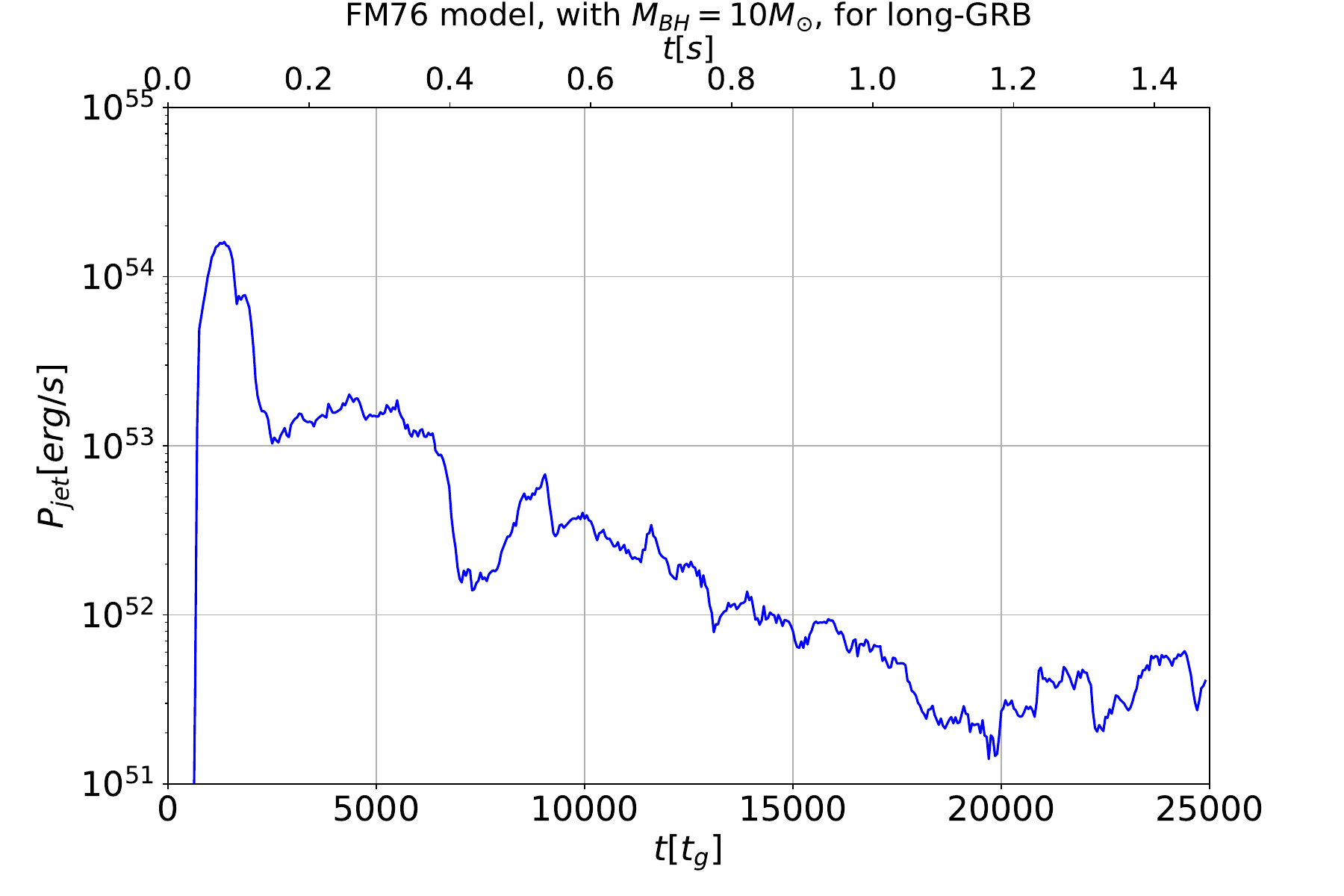}
\includegraphics[width=0.47\textwidth]{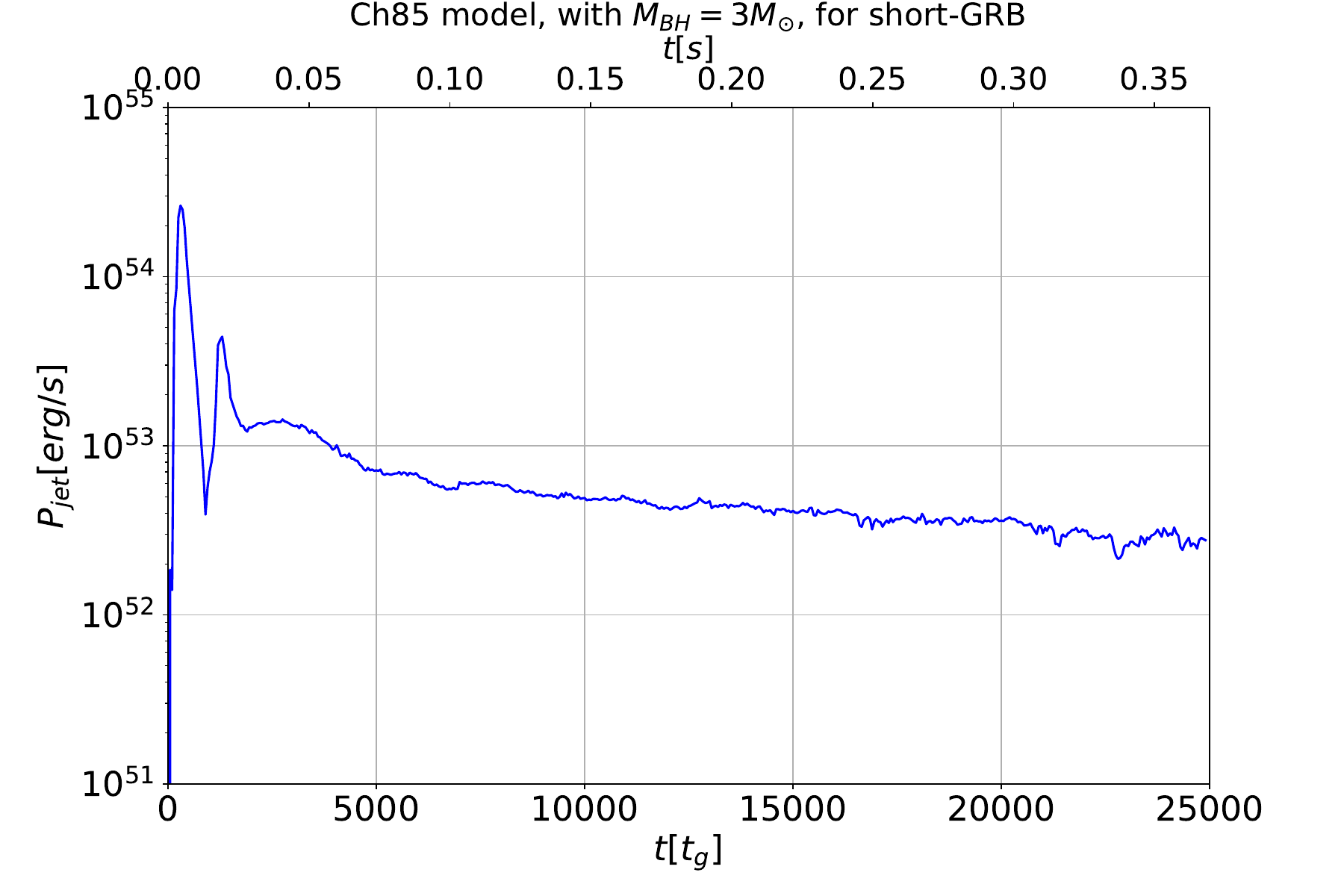}
\caption{Jet power estimated by the Blandford-Znajek luminosity (see Equation \ref{eq:p_jet}) in physical units (erg/s) as a function of time (in $t_g$). We also show the physical timescales scaled with the assumed black hole masses of long and short GRBs: (a) with initial Fishbone-Moncrief configuration (FM76) and (b) with initial Chakrabarti configuration (Ch85).
}
\label{jet_pow_t}
\end{figure*}

We show the evolution of the jet power with time estimated from the Blandford-Znajek luminosity in Figure \ref{jet_pow_t}. Notice that the plots are also marked with physical units assuming a central black mass of 3 $M_{\odot}$ for the short GRB central engine and a mass of 10 $M_{\odot}$ for the long GRB progenitor central engine (see \cite{Sharmaetal2021BHmass} for an estimated range of black hole masses for observed GRB samples). It can be noticed from the plots that the jet power rises after a certain amount of time after the simulations have begun; the jet formation time is actually coincident with the development of the toroidal magnetic field component in the disk as described in section \ref{torus_evolution}. 

\subsubsection{Variability of the jet and accretion rate}

\begin{table*}[htb]
\centering

\begin{tabular}{@{}lcccccccc@{}}
\toprule
\multirow{2}{*}{Model} & \multicolumn{3}{c}{Lorentz factor ($\Gamma$)} & \multicolumn{3}{c}{MTS estimated (in $t_g$)} & \multicolumn{2}{c}{Slope of the PDS} \\ \cmidrule(l){2-9} 
& Location 1 & Location 2 & Average & Location 1 & Location 2 & Average & Location 1 & Location 2 \\ \midrule
FM76 & 105.96      & 89.45      & 97.71   & 178.63     & 269.80     & 224.21  & -0.8253   & -1.4899    \\
Ch85 & 61.33      & 202.36      & 131.85   & 147.01     & 147.72     & 147.37  & -0.8016   & -1.1310    \\ \bottomrule
\end{tabular}
\caption{The values of the bulk Lorentz factor ($\Gamma$), estimated minimum variability timescale (MTS) (in $t_g$) and the slopes of the power law (PL) fit to the power density spectrum (PDS) plots, at the two chosen locations inside the jet for our models.}
\label{table:time_variability}
\end{table*}

\begin{figure*}
\centering
\includegraphics[width=0.47\textwidth]{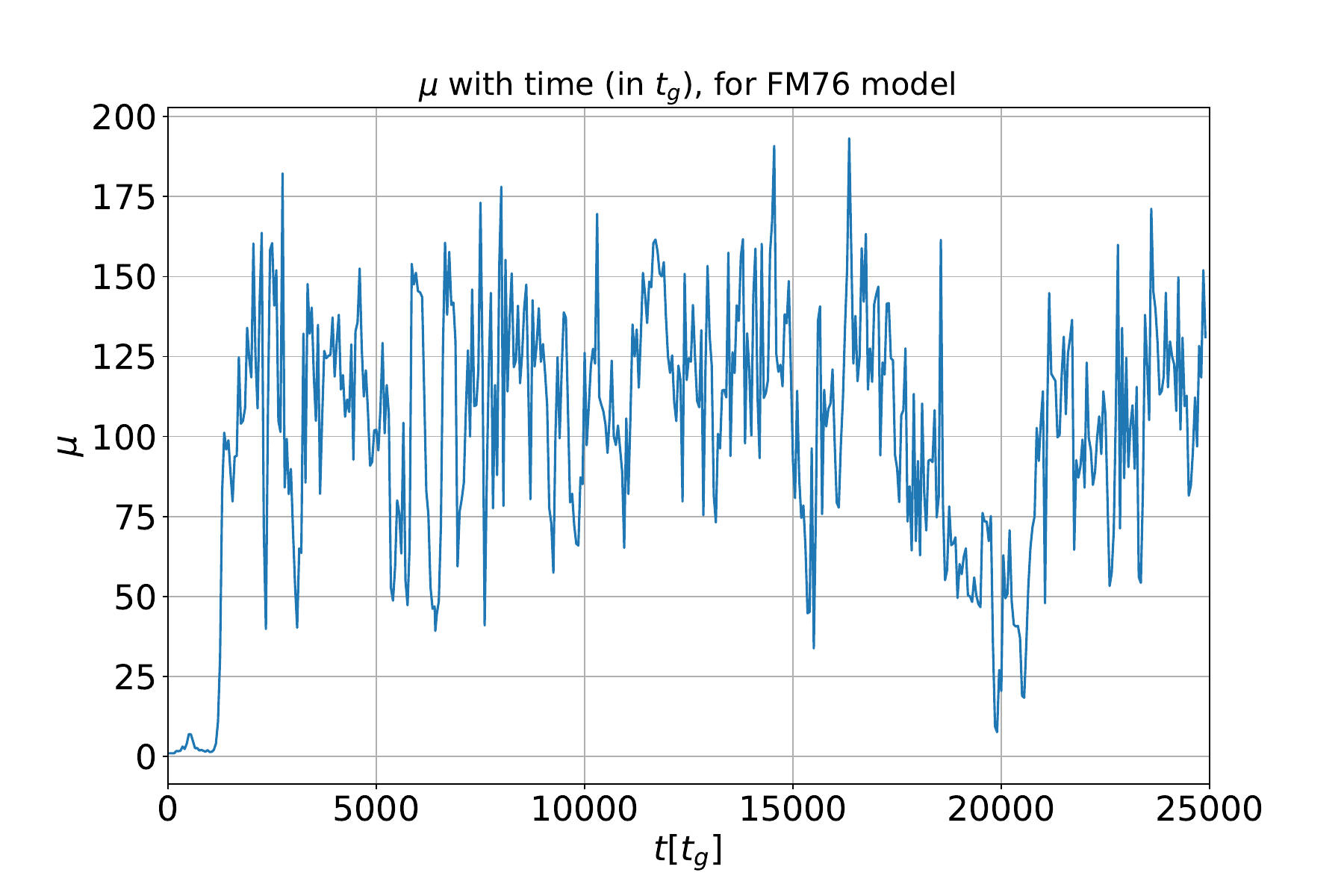}
\includegraphics[width=0.47\textwidth]{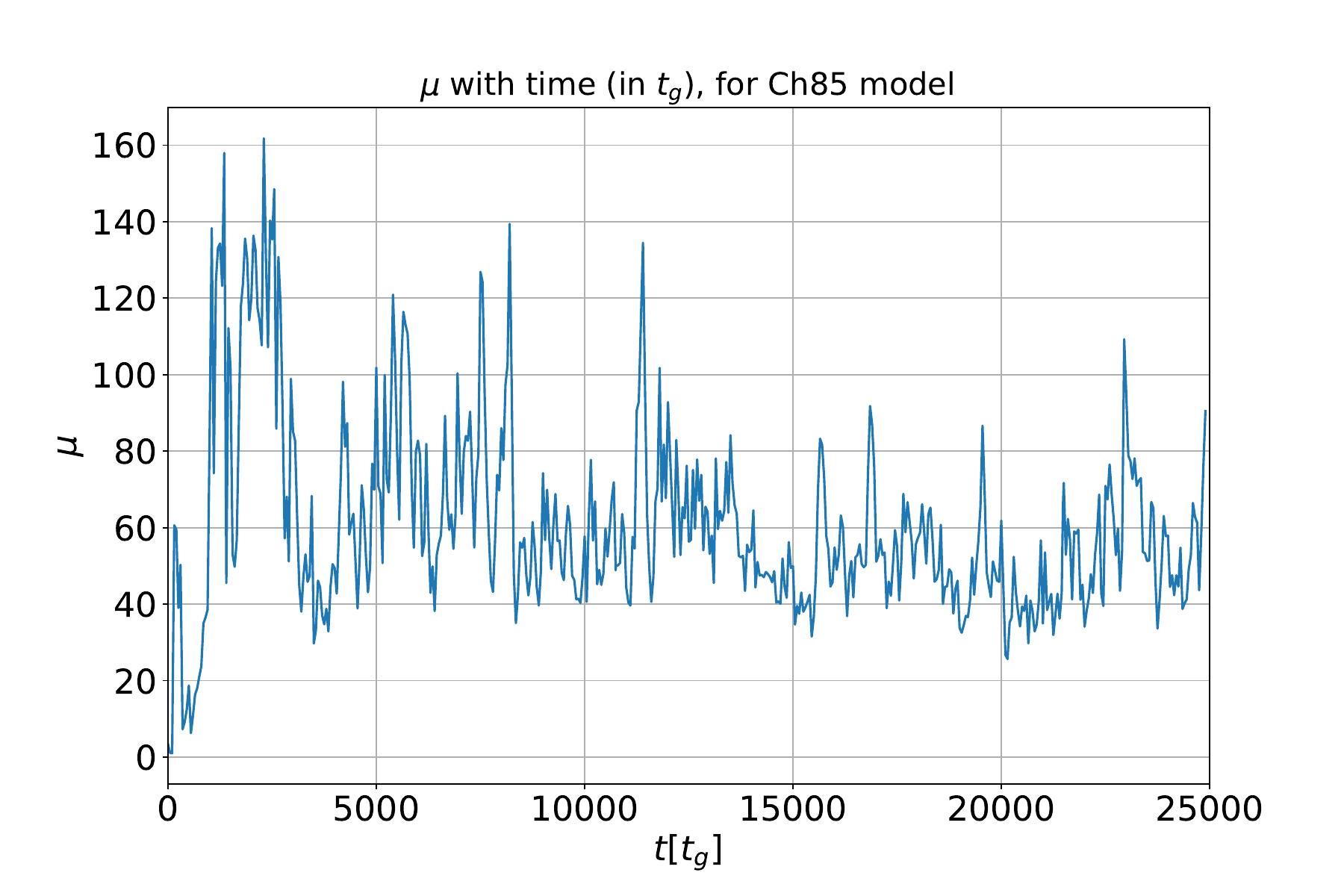}
\caption{The time evolution of the jet energetics parameter $\mu$ at a chosen location $r  = 150 r_g$ and $\theta = 5^{\circ}$ (location 1), averaged over the toroidal angle $\phi$ for the models (a) with initial Fishbone-Moncrief configuration (FM76) and (b) with initial Chakrabarti configuration (Ch85).
}
\label{fig:mu_t_p1}
\end{figure*}

\begin{figure*}
\centering
\includegraphics[width=0.47\textwidth]{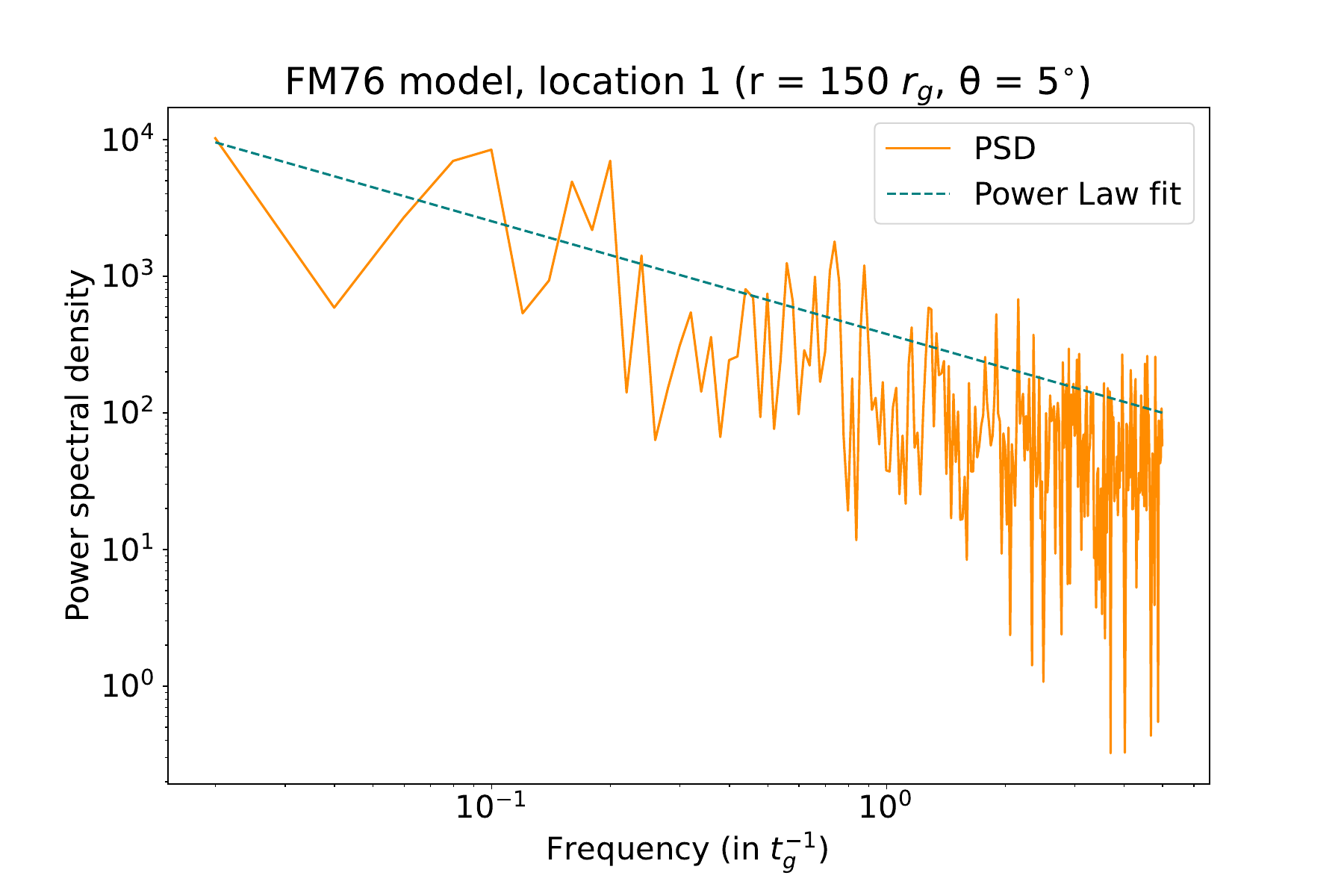}
\includegraphics[width=0.47\textwidth]{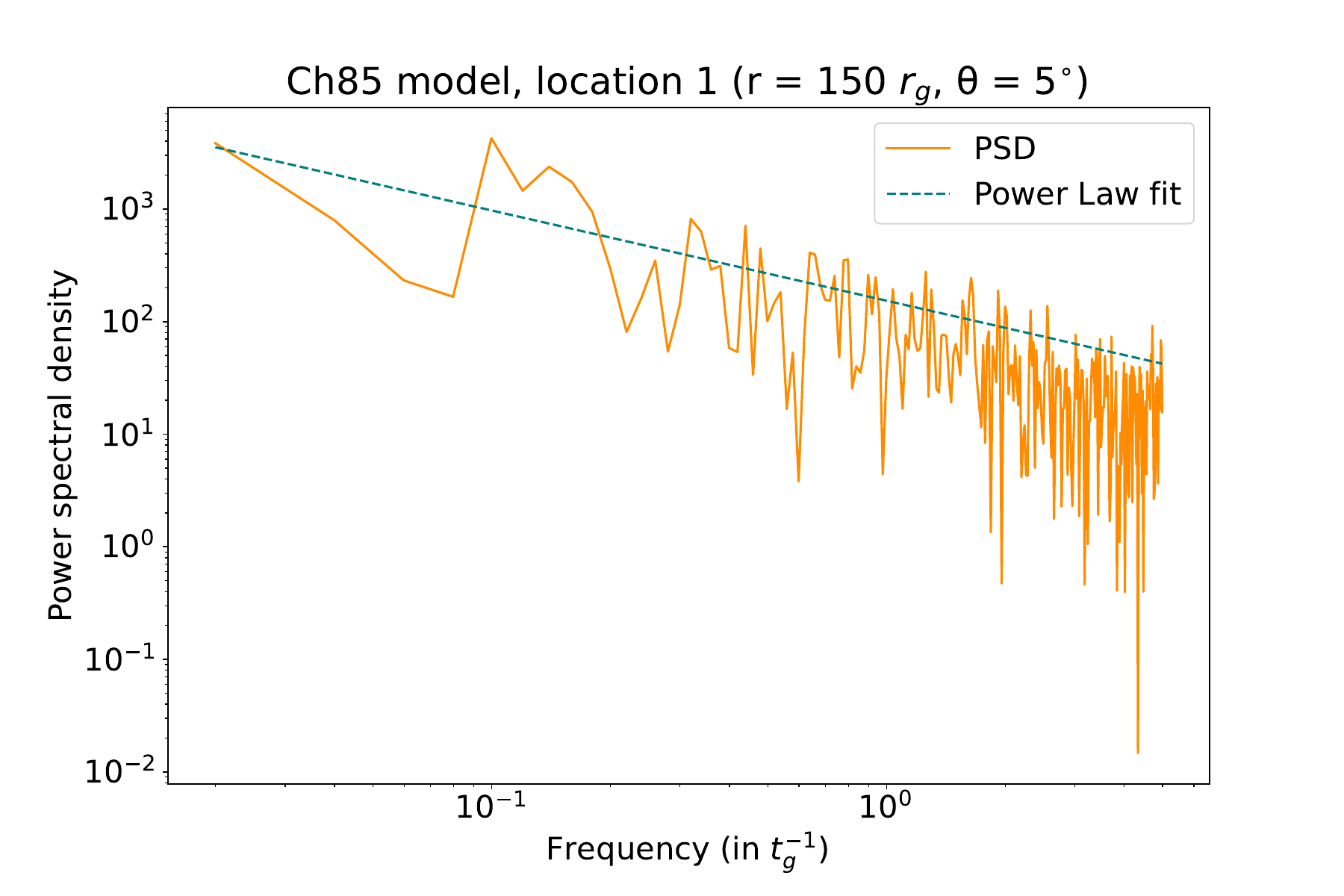}
\includegraphics[width=0.47\textwidth]{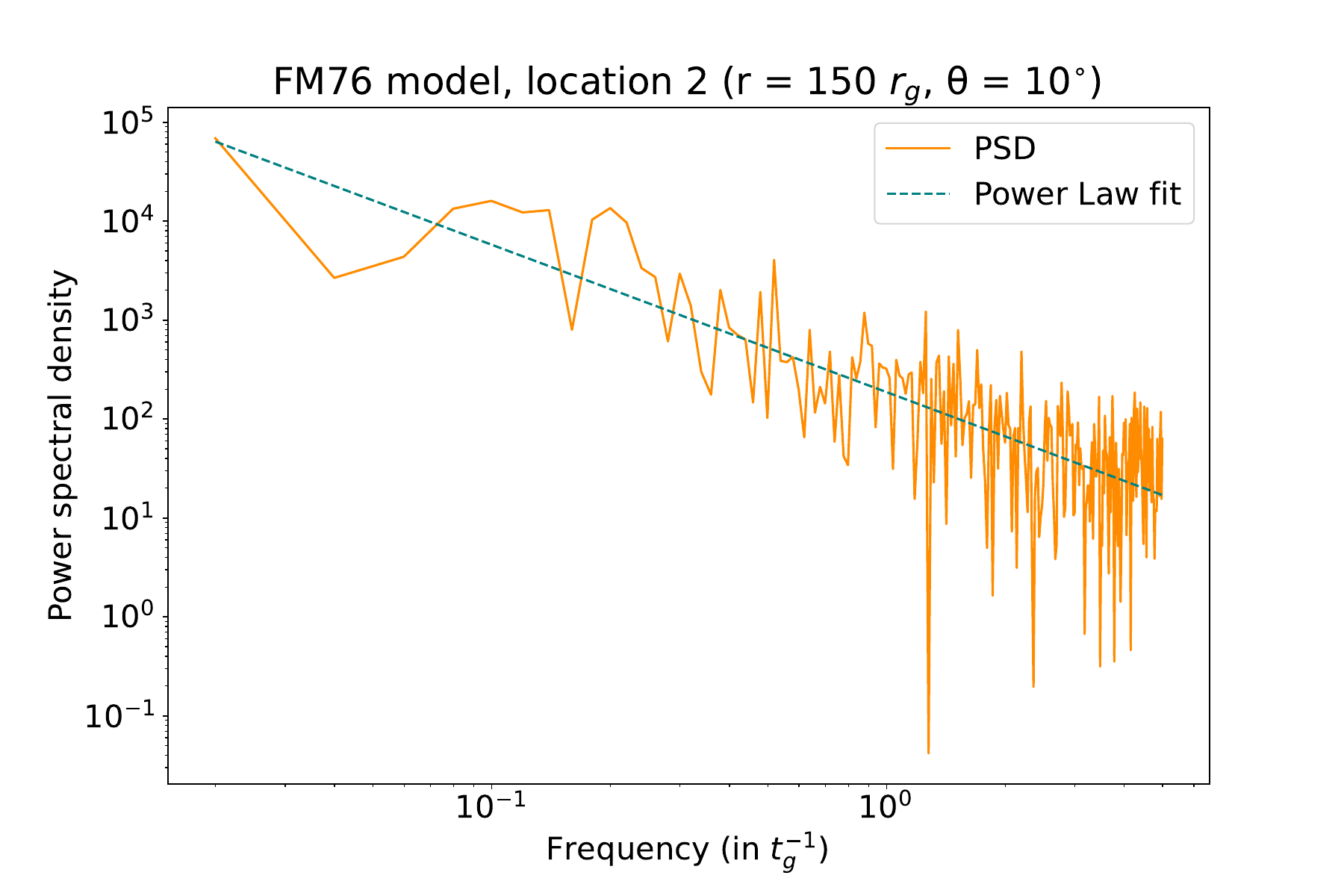}
\includegraphics[width=0.47\textwidth]{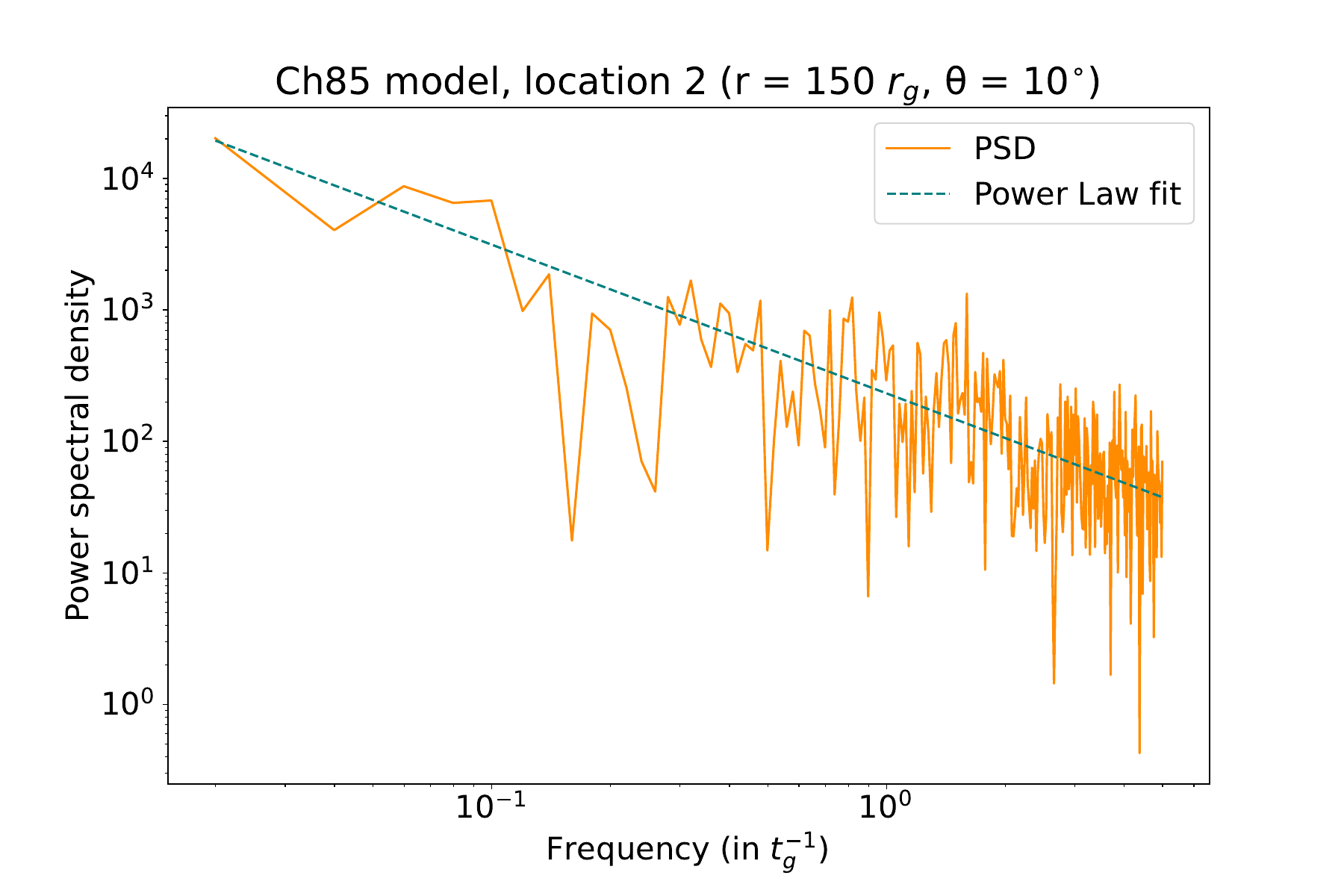}
\caption{The power spectral density computed from the $\mu$-time variability data at the chosen locations 1 and 2 and the  corresponding power law fit for the models (a) with initial Fishbone-Moncrief configuration (FM76) and (b) with initial Chakrabarti configuration (Ch85).
}
\label{fig:psd_loc1}
\end{figure*}

Figure \ref{fig:mu_t_p1} shows the variability of the jet energetics parameter with time as measured at the location 1, that is at the inner region closer to the axis of the jet, for our two models. We calculate the average duration of the peak widths at their half maximum and use it as a proxy for the minimum variability time scale (MTS) of the jet. The MTS computed from the above chosen locations and their averages are given in Table \ref{table:time_variability}. The variability time scale has a smaller value in the inner region of the jet for both the models and thus the peaks have shorter duration at smaller angles from the rotation axis of the black hole. To better understand the time variability of the jet emission we also computed the power density spectra (PDS) of the $\mu$ variability data from these chosen locations. We fit this data to a power law function of the form $y(x) = A x^{\alpha}$ and calculate its slope. Figure \ref{fig:psd_loc1} shows the PSD and the power law fit for the $\mu$ data from both the locations, for our models. The slopes of the fitted power law curves are also given in Table \ref{table:time_variability}.

\subsubsection{Jet profile}

Figure \ref{fig_mu_2d_fm} shows the 3D distribution of the jet energetics parameter $\mu$ for both of our models up to a radius of 200$r_g$ at a dynamical time 5000$t_g$. These plots represent the evolved structure of the jet after a certain time after the jet has launched. 
Both of our models clearly produce structured jets rather than a simple top-hat, similar to what we have seen in our previous study with 2D simulations. But our current models evolve the non-axisymmetric structure of the jet. Larger values of Lorentz factor are obtained farther from the axis as can be seen from the plots. So the jets produced by our models have a comparatively hollow core, with small Lorentz factors ($\sim 10$) up to an angle of $\sim 5^{\circ}$. Higher Lorentz factors are reached far from the axis the jet and it is faster along the edges. This is similar to those observed in previous simulations of MHD jets as described in \cite{Nathanail2020openangle}. Further description of the jet structure is provided in the next section where we discuss it in the context of short GRBs.

\begin{figure*}
\centering
\includegraphics[width=0.47\textwidth]{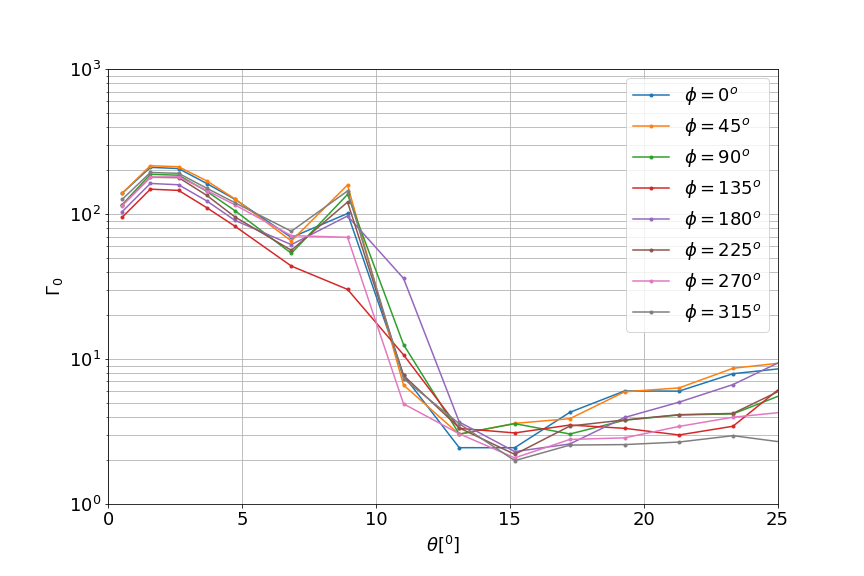}
\includegraphics[width=0.47\textwidth]{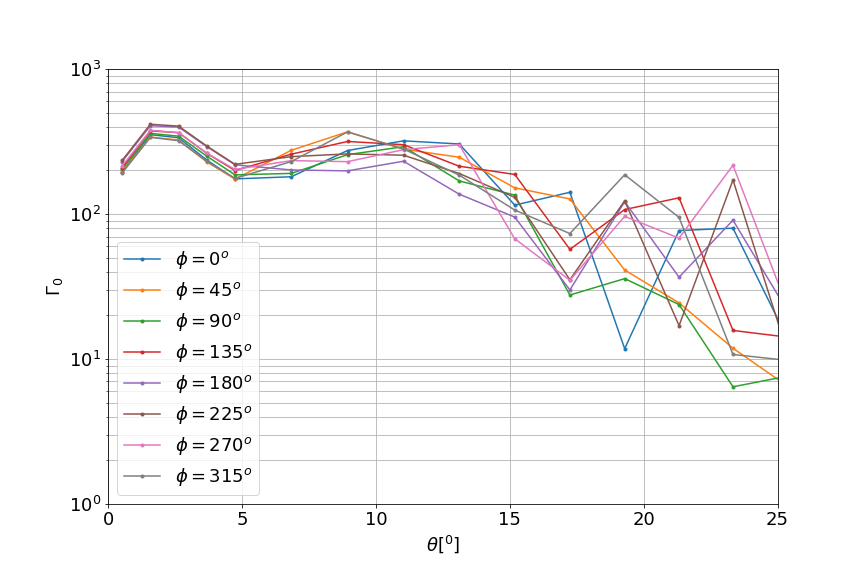}
\caption{Time averaged jet Lorentz factor measured at a large distance of 2000 $r_g$, as a function of the polar angle $\theta$. Figure shows the jet profile at different selected slices of $\phi$ for the model with (a) Fishbone-Moncrief initial configuration (FM76) and (b) Chakrabarti initial configuration (Ch85).
}
\label{fm_profile}
\end{figure*}

In order to qualitatively understand the energy distribution in the jet, we calculated the jet Lorentz factor as a function of the polar angle at a very large distance from the center. Figure \ref{fm_profile} shows the time averaged jet Lorentz factor estimated at a large distance (2000$r_g$) as function of the polar angle for both the models. For the FM model, the most energetic parts of the jet is confined mostly within an angle of $\sim 11-12^{\circ}$ and the highest Lorentz factors are reached around $9^{\circ}$ from the axis. We discuss this model in the context of long GRBs. On the other hand, the model with Chakrabarti initial configuration, which we consider as the central engine for short GRBs, has a somewhat different jet profile. In this model, the jet does not clearly confine to an angle of $\lesssim 15^{\circ}$ but instead spreads up to $25^{\circ}$. It is worth noting that the inner part near to the axis of rotation is much less energetic in this model. The outer regions of the jet, which is away from the rotational axis of the black hole, are also more structured in this model. The value of the Lorentz factor along different $\phi$ slices vary significantly between each other.

\subsection{Effect of the resolution and the initial magnetic field geometry in achieving the MAD state}

The resolution study of \cite{White2019} was particularly focusing on the MAD scenario. They find that certain global properties converge with resolution such as the mass accretion rate, jet efficiency and the MRI suppression factor. We note that our global resolution of $288\times256\times128$ is at least at the level of their level 2 resolution which satisfactorily resolves these quantities. 

\begin{figure}
    \centering
    \includegraphics[width=0.48\textwidth]{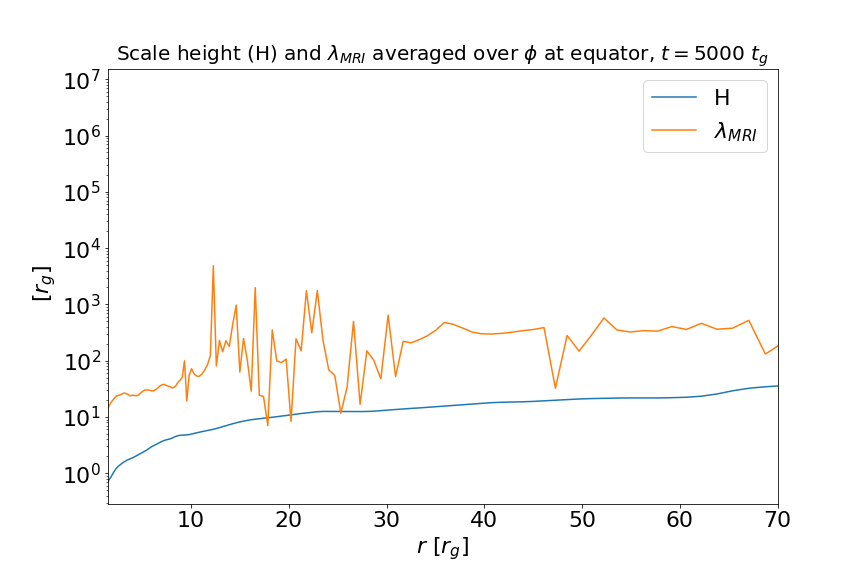}
    \includegraphics[width=0.48\textwidth]{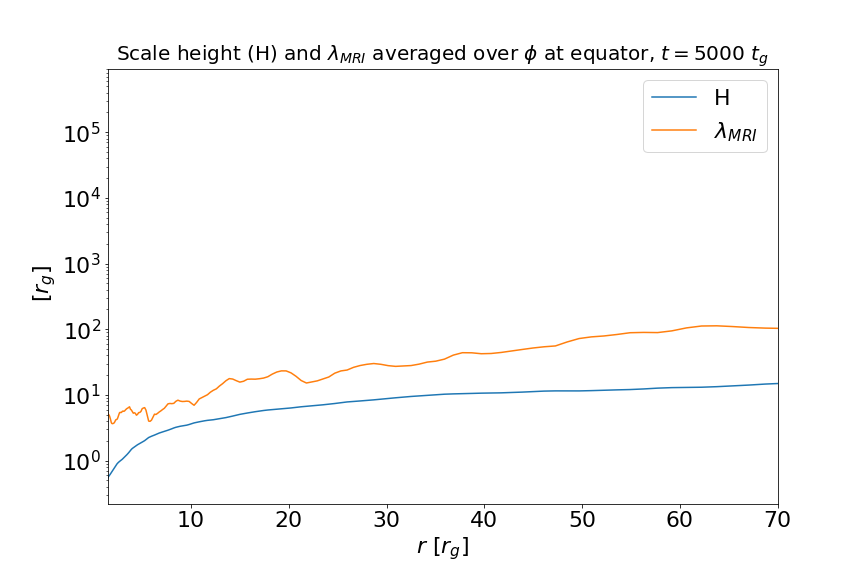}
    \caption{Comparison of the disk scale height (H) and the MRI wavelength at the equator, averaged over the azimuthal angle $\phi$, at an evolved time $t = 5000 ~t_g$ for (a) FM76 model and (b) Ch85 model. We have $\lambda_{MRI} \geq H$ throughout most part of the disk in both cases, indicating that the magnetic field is strong enough to suppress the the MRI.}
    \label{fig:H+lambda_MRI}
\end{figure}

The wavelength of the fastest growing mode of the MRI is given by \citep{siegel2018LambdaMRI}:
\begin{equation}
    \lambda_{MRI} = \frac{2\pi}{\Omega}\frac{b}{\sqrt{4\pi \rho h + b^2}}
\end{equation}

Initially, accretion is initiated and driven by the MRI as described previously. 
The MRI is then suppressed in our models at an evolved time by the fact that a very strong magnetic barrier is formed near the black hole horizon. This halts the MRI driven accretion,
and interchange instabilities drive the process,
resulting in the formation of a MAD state.
The suppression of MRI can be quantified by the parameter $S_{MRI}$, which is the ratio of the MRI fastest growing mode wavelength over the disk scale height (cf. \cite{White2019} \& \cite{McKinney2012MagChoked}):
\begin{equation}
    S_{MRI} = \frac{2H}{\lambda_{MRI}}
\end{equation}

where $H=c_s / \Omega_K$ with the sound speed $c_s = \sqrt{\gamma p_g / (\rho_0+u_g+p_g)}$ and $\Omega_K$ is the Keplerian angular velocity.
Figure \ref{fig:H+lambda_MRI} shows the comparison of the disk scale height and the MRI wavelength in our models at an evolved time $t = 5000 ~t_g$. We can see that the MRI wavelength is much larger in comparison to the disk scale height in both cases indicating that $S_{MRI} < 1$ and the suppression of the MRI.

Following \cite{White2019} we assume that the accumulated flux on the horizon $\phi_{BH}$ is not affected by the resolution. Thus we argue that the large scale structures of the MAD flows are correctly described by our models. Our models have sufficiently enough resolution to properly catch the base of the jet structure and Lorentz factor profiles as well. 

We initialize our models with single large poloidal loops of magnetic field. Such configurations are proved to result in the development of strong poloidal flux in the vicinity of the black horizon as the accretion proceeds \citep{McKinney2012MagChoked, Penna2013membrane}. This helps in achieving the MAD state within a short time ($\sim 1500 - 2000 t_g$) after the simulation has begun (as can be seen from Fig.\ref{fig:phi_BH}). We note, that some recent simulations have shown that the development of such strong poloidal flux happens even with much weaker initial fields \citep{Liska2020} and also develops a magnetically arrested accretion disk state. In order to see such effects, we would need much higher resolutions and longer runs as compared to the ones presented in this paper. However, some of our models, previously run in 2D, and for shorter amounts of time with weaker initial magnetic field configurations, also showed the building up of strong poloidal flux after a certain time 
\citep{Janiuketal2021}.

\section{Application to the engine modeling of short and long GRBs}\label{applications}

\subsection{Comparison with previous models for GRBs} 

We use our Ch85 model in the context of short GRBs, as a post-merger system in which an accretion disk has been formed. In a realistic scenario of a BNS system, the disrupted neutron star material is  non-axisymmetic and composed of clumps of matter, with  propagating shock waves \citep{Foucart2014}. Therefore, breaking the initial axisymmetry of the disk with the introduction of random perturbations in the $\phi$-direction is justified. 

 \cite{Proga2006} used their hyperaccretion models for GRBs at late times and proposed that the episodic energy output in the jets is connected with the changes in the mass supply driven by the accretion rate. They propose this as a model to explain the X-ray flares observed from the GRBs. They show with their chosen models that the energy release is repeatedly halted and restarted given the mass supply rate decreases with time which is the case in both binary merger and collapsar scenarios. This is similar to behaviour observed in our previous work \cite{Janiuketal2021}. 
In our current models there is no mass supply to the outer edge of the disk.
The short episodic rise in the mass accretion rate in the initial stages is due to the dynamic nature of the magnetic field, and relaxation of the initial, stationary conditions, which was derived for a non-magnetized disk. 
After the initial condition is relaxed, i.e. around time $5000 t_g$, the variability of accretion rate driven by interchange instabilities developed in the plasma continues in the magnetically arrested disk. 
The jet power is also varying and its magnitude is slowly decreasing with time. This can be partly connected to the falling mass accretion rate, which is decreasing at a slightly different rate. The rate of decrease of the jet power in both the models differs substantially. In the FM76 model it falls by three orders of magnitude as the simulation proceeds while in the Ch85 model it falls by two orders of magnitude. The differences can be connected to the varying strength of magnetic flux on the black hole horizon.  

\cite{Lloyd-Ronning2019a} put limits on the magnetic field strength and the black hole mass needed to power Blandford-Znajek jets from the observed luminosities of short and long GRBs. They assume a central engine with black hole mass in the range of 0.5-4 $M_{\odot}$ for the short GRBs and 2-10 $M_{\odot}$ for the long GRBs and they find that magnetic field strengths in the range of $ \sim 5 \times 10^{14}$ to up to $\sim 10^{17} G$ (for the long GRBs) and $\sim 10^{15}$ to $\sim 10^{17} G$ (for the short GRBs) are needed to power the observed GRBs with the assumed mass of the central engine. The magnetic fields inferred from their analysis are extreme and in practice such fields can be generated and sustained only through special mechanisms. 
Generally, a small extant magnetic field in a disk can be amplified by the MRI. But, the MRI is suppressed in our simulations due to the magnetic arresting of the disk which is evident from the growth of magnetic flux at the black hole horizon. The magnetic field strength at the black hole horizon at an evolved time (5000$t_g$) in our long GRB model (FM76) is estimated as $3.51 \times 10^{14} G$ and in our short GRB model (Ch85) it is $7.74 \times 10^{14} G$. This is in the range of previous estimations. Our simulations show that the disk can sustain such an amount of flux near the black hole horizon over time and presents such a configuration as a viable candidate for explaining the jet variability properties of the GRBs. For comparison, \cite{Kiuchi-2014} evolve a high resolution BNS merger scenario. In their simulations magnetic field is amplified within multiple mechanisms including Kelvin-Helmholtz instability during the merger and MRI during post-merger evolution. They show it is possible to have a highly magnetized disk with magnetic field stronger than $10^{15.6}$G over a big region of the disk. 

Our FM76 model (model 1) is embedded in a poloidal magnetic field which follows the disk density structure with a dependence on the fifth power of radius. This analytic solution is a standard initial condition for a thick disk and the imposed magnetic field is in such a way that it takes a short amount of time to accrete more poloidal flux to the black hole horizon. We consider this model as a candidate for the central engine of long GRBs. On the other hand, we consider our Ch85 model (model 2) embededd in a poloidal magnetic field due to a circular current at the radius of pressure maximum of the disk as a candidate for the central engine of short GRBs. In this model, the imposed magnetic field is in such a way that it takes shorter time to develop more poloidal magnetic flux on the black horizon, which results in a magnetically arrested disk. At a later time of the simulation ($t \sim 2500 t_g$), the strength of the poloidal magnetic field at the inner region of the disk is $B_z = 1.53 \times 10^{-3} $ for the FM76 model and $B_z = 1.09 \times 10^{-3} $ for Ch85 model (in code units) which is comparable.

\subsection{Jet opening angles in short and long GRBs}

Our numerical simulations can give a picture of the jet structure in the long and short GRBs based on the models we are considering for the two scenarios. 
The time averaged jet Lorentz factor profile with $\theta$ gives information about where most of the energy in the jet is concentrated and the jet opening angles in our models. 
Based on their HD and MHD simulations \cite{Nathanail2020openangle}
focus particularly on the properties of GRB170817A. They observe a self consistent jet launching in their MHD simulations. Their MHD jets have a hollow core of up to $\sim 4-5^{\circ}$ and the jets carry a significant amount of matter. When comparing to our models, we have a clearly hollow core up to an angle of $~5^{\circ}$ in our Ch85 model. On the other hand, the jet launched in our FM76 model has higher Lorentz factors closer to the axis as well, compared to the Ch85 model. 
But, in our FM76 model also, even higher Lorentz factors are reached at a significant angle ($\theta \gtrsim 10^{\circ}$) from the axis. Compared to their models, the jets in both of our models do not contain any considerable amount of matter. This is apparent from the density plots on the top middle and right panels given in Figures \ref{rho_B_fm} and \ref{rho_B_ch}. 

\cite{Margutti2018openangle} studied different models based on hydrodynamic simulations of the jet interaction with BNS ejecta to get the opening angle values from the GRB170817A afterglow observations. 
They show two representative scenarios for jet opening angles of $5^{\circ}$ and $15^{\circ}$ considering a top-hat structure for the jet, assuming off-axis viewing angles $\theta_{obs} \sim 15^{\circ} - 25^{\circ}$. But the top-hat jets viewed off-axis failed to reproduce the larger X-ray and radio luminosities in the early days after the prompt emission from the source and also failed to account for the mild and steady rise of the non-thermal emission observed. So a simple top-hat structure is not a likely scenario for the afterglows of GW170817. Rather, a structured and collimated relativistic outflow is considered as the probable scenario for this observation, with the GRB assumed to be seen off-axis. Many other authors also support the scenario of a structured jet which is seen off-axis based on their models for this source. e.g. \citep{Kathirgamaraju2018offaxis}. 
If we assume the inference of \cite{Margutti2018openangle} that the GRB170817A was a "classical" short GRB with a structured collimated jet and reached up to an energy of $\sim 10^{50}$ erg which was observed off-axis to be correct, we can consider the jet structure described by their models. The jet produced in their representative models with this assumption, has a narrow ultra-relativistic core of $\theta_c \sim 9^{\circ}$ with $\Gamma \sim 100$ surrounded by a mildly relativistic sheath up to angle of $\lesssim 60^{\circ}$. 
In our models, the time averaged terminal Lorentz factor profile computed at $\sim 2000 r_g$ (shown in Figure \ref{fm_profile}) also clearly shows the existence of structured jets. 
The Ch85 model, which we consider as a candidate for a short GRB central engine, also shows a complex structured jet profile with no clearly defined core region when averaged over time. 
In this model, the jet is confined to a rather broader angle of $\theta \lesssim 25^{\circ}$.
But, the jet produced in this model reaches the highest Lorentz factors at around an angle of $\sim 9^{\circ}$ which is in coincidence with the value of narrow jet core in the models of \cite{Margutti2018openangle}. 
Figure \ref{fig_mu_2d_fm} shows the internal sub-structure of the base of the jet from our simulations at a given time. Both these plots show the evolved non-axisymmetric structured nature of the jet. 
They also qualitatively depict the jet core with higher Lorentz factors ($\sim 150 - 200$) at some certain angle away from the axis.

Another possible scenario, explaining why GRB 170817 is dimmer than a classical short GRB, is based on the fact that post-merger remnants are surrounded by high-speed, neutron-rich dynamical ejecta which produce heavy elements with high opacity through r-process nucleosynthesis. Therefore the emissions from inner parts can be partially masked by outer opaque material ~\cite{Kasen-2015}. GRB 080503 is another example of a faint GRB, accompanied by an extremely bright extended prompt X-ray emission. The faintness of this GRB can be explained by off-axis jet similar to GRB170817~\citep{Perley_2009, Fongetal2015}. 


\cite{Fongetal2015} considered 11 short GRB events with equal weighting (see their Table 5) and calculated the jet opening angles by giving Gaussian probability distributions to the measurements they considered. Employing a realistic upper bound of $\theta_{j,max} = 30^{\circ}$ on the opening angles (obtained from the previously available post-merger black hole accretion simulations), they obtain a median value of $\langle \theta_{j} \rangle = 16 \pm 10$ degrees. The jet from our Ch85 model has a value which tend towards the upper limit of this value. 
In the FM76 model we have a structured jet with most of the energetic part confined to $\theta \sim 11-12^{\circ}$. The profile for this model shows an ultra-relativistic core with $\Gamma \sim 100$ peaking around $\theta_{c} \sim 9^{\circ}$. 
\cite{Fongetal2015} also compute the jet opening angles for the long GRBs from the measurements of 248 samples and find a median value of $\langle \theta_{j} \rangle = 13_{-9}^{+5}$ degrees. The value of $11^{\circ}$ obtained from our FM76 model matches with the lower bound of the estimated value from the observations. 

\subsection{Jet variability timescale in short and long GRBs}

When we consider our model with Chakrabarti initial torus (Ch85) as a plausible central engine for the short GRBs, we can use a black hole mass of 3 $M_{\odot}$ to scale the physical quantities and compute the minimum variability time in seconds. We get an average value of 147.37 $t_g$ for the minimum variability timescale from two chosen locations along the jet direction for this model. This scales to a value of 2.178 ms in physical units. Similarly when we consider our FM76 model as a collapsar remnant disk and use a value of 10 $M_{\odot}$ for the central black hole, we can estimate the resulting minimum variability timescale. We get an average value of 224.21 $t_g$ for the minimum variability timescale in code units from this model which gives a value of 13.257 ms in physical units.  
\cite{MacLachlan2013a} present a wavelet analysis using a collection of observed samples from the FERMI data to show that the variability timescales of the short and long GRBs differ from each other. In their study, they find a marginal positive correlation between the minimum variability timescale and burst duration for both classes of bursts. The lowest variability timescale from their study is 3 ms. 
Our results are in agreement with these observations.

\cite{Golkhou2015} put constraints on the minimum variability timescales for the GRBs observed by the FERMI/GBM instrument. They find the median minimum timescale for long GRBs as 45 ms and for the short GRBs as 10 ms. Only less than 10\% of their selected samples show variability smaller than 2 ms and this requires Lorentz factors higher than 400. 
The minimum variability timescales obtained from our models are in the lowest range of these observed values. 

The power density spectra (PDS) computed for our models also reveal information about the time variabilty data. The variability of the energy blobs ejected close to the inner regions of the jet in both models have a smaller PDS slope as compared to the variability measured at the outer region for the fitted power law. So, in general, the outer wall of the jet shows higher variability as compared to the inner part closer to the jet axis. Also, the PL fit for the long GRB model has steeper slope as compared to the short GRB model at both the locations in which the variability is measured. 
The slope ($\alpha$) values are in the range but smaller than 5/3 which is predicted by the turbulence model \citep{Beloborodov2000}. More recent observational studies give PDS slope values, with a PL fit, in the range 1.347 to 2.874 \citep{Guidorzi2016lGRBpds} and 1.42 to 4.95 \citep{Dichiara2016pds} for the long GRBs and in the range 1.398 to 2.507 \citep{Dichiara2013b-SGRBpds} for the short GRBs. Our results of the jet variability measured at the outer part are consistent with the lower limit of the observed PDS slopes.

\cite{Sonbas2015} find an anti-correlation between the minimum variability timescale and the bulk Lorentz factor for the GRB samples they considered from the SWIFT and FERMI data. We have not calculated sufficient number of models to do a parameter study for obtaining a correlation between the variability timescale and Lorentz factor.
But our models do systematically give smaller minimum variability timescales for higher Lorentz factors from our two models and also from different regions of the jet.

\section{Summary and Conclusions}\label{conclusions}
We have done 3-dimensional GRMHD simulations of the accretion disk around a Kerr black hole considering two different initial analytical equilibrium solutions. The Chakrabarti initial solution has an angular momentum distribution constant over specific surfaces and changes with the radius rather than being constant over the entire disk. This arguably gives a more realistic scenario as compared to the initial FM solution. These solutions are imposed upon by poloidal magnetic fields which results in magnetic turbulence and MRI which initially drives the accretion. As the accretion proceeds the plasma brings more magnetic flux to the black hole horizon and results in a magnetic arrested accretion state which prevents further falling of matter into the black hole. We consider the MAD state as the plausible central engine of GRBs and investigate its effect on the time variability properties and the structure of the ejected jet. Our models self consistently produce structured jets with a relatively hollow core, up to an angle of $\sim 5^{\circ}$, and higher Lorentz factors are reached on the edges of the jets. The two initial models and the imposed magnetic field geometry affect the jet structure which is evident from the different profiles and internal structures of the jets. The models give a jet opening angle of $\sim 11^{\circ}$ and $\sim 25^{\circ}$ respectively for the short and long GRBs, which is consistent with the observations. Also the minimum variability timescale is computed for both of our models and they are found to be in the lower range of variability timescale from observations. Finally, the PDS spectral slopes are somewhat flatter than the classical $\alpha = 5/3$ slope, but within the observed range.

From our work, it can be seen that the MAD model can serve as a plausible central engine in the context of both the classes of GRBs. In the collapsar scenario, our models can be considered as an initial approximation, as we do not evolve it for longer timescales observed for such systems. In the short GRB regime, our models show that a MAD can be formed in such short timescales if the initial magnetic fields are strong enough near to the black hole horizon.  The structured jets and the opening angles from our models are in agreement with the values from the observations of a collection of GRB samples in both the cases. However, jet structure and opening angles vary significantly between each individual burst and such differences need further investigation of more possible initial configurations. Our models are also able to explain the sub-second time variability and the MTS values are in good agreement with the observations. The observed correlations between the MTS and bulk Lorentz factor of the jets can be investigated by a further parameter study of our models with different initial magnetic field strengths and black hole spin values. This can also help in understanding the dependence of the jet structure and opening angles on the varying initial conditions. There is also possibility of more detailed studies of time variability from our models to detect coherent fluctuations at various radii and propagating signals or time lags. (e.g. as in \cite{Bollimpalli2020})

\begin{acknowledgements}
We thank the anonymous referee for the valuable comments that helped to improve our manuscript. We thank Kostas Sapountzis and Om Sharan Salafia for helpful discussions. This research was supported in part by the grant DEC-2019/35/B/ST9/04000 from the Polish National Science Center. This research was carried out with the support of the Interdisciplinary Center for Mathematical and Computational Modeling at the University of Warsaw (ICM UW) under the grant numbers g85-986 and g86-987.

\textit{Software:} \texttt{Matplotlib} \citep{Matplotlib2007}, \texttt{NumPy} \citep{Numpy_article2020}, \texttt{SciPy} \citep{SciPy2020} and \texttt{VisIt} \citep{VisIt2012}.
\end{acknowledgements}

\bibliography{ref.bib}

\end{document}